\documentclass[reprint,superscriptaddress,preprintnumbers,showpacs]{revtex4-1}
\usepackage{graphicx}
\usepackage{dcolumn}
\usepackage{bm}
\usepackage{multirow}

\newcommand{\eq}[1]{eq.~(\ref{#1})}

\newcommand{\Eq}[1]{Eq.~(\ref{#1})}
\newcommand{\Eqs}[2]{Eqs.(\ref{#1},\ref{#2})}

\newcommand{\ur}[1]{(\ref{#1})}
\newcommand{\urs}[2]{(\ref{#1},\ref{#2})}

\newcommand{\beq}{\begin{equation}}
\newcommand{\eeq}{\end{equation}}

\newcommand{\la}[1]{\label{#1}}
\newcommand{\bea}{\begin{eqnarray}}
\newcommand{\eea}{\end{eqnarray}}
\newcommand{\ba}{\begin{array}}
\newcommand{\ea}{\end{array}}

\newcommand{\half}{{\textstyle{\frac{1}{2}}}}

\newcommand{\noi}{\noindent}
\newcommand{\n}{\nonumber}
\newcommand{\nn}{\nonumber}
\newcommand{\Tr}{{\rm Tr}}
\newcommand{\Sp}{{\rm Sp}}

\newcommand{\tOmega}{\tilde{\Omega}}
\newcommand{\tomega}{\tilde{\omega}}
\newcommand{\exc}{\text{excited}}
\newcommand{\K}{{\cal K}}

\begin{document}

\title{  A theory of baryon resonances at large $N_c$}
\author{\framebox{Dmitri Diakonov}$\;$}
\affiliation{Petersburg Nuclear Physics Institute, Gatchina 188300, St. Petersburg, Russia}
\affiliation{Ruhr-Universit\"at Bochum, Bochum D-44780, Germany}
\author{Victor Petrov}
\affiliation{Petersburg Nuclear Physics Institute, Gatchina 188300, St. Petersburg, Russia}
\affiliation{Ruhr-Universit\"at Bochum, Bochum D-44780, Germany}
\author{Alexey A. Vladimirov}
\affiliation{Ruhr-Universit\"at Bochum, Bochum D-44780, Germany}
\affiliation{Department of Astronomy and Theoretical Physics, Lund
University, S 223 62 Lund, Sweden}

\preprint{LU-TP 13-31}

\begin{abstract}
At large number of colors, $N_c$ quarks in baryons are in a mean field of definite
space and flavor symmetry. We write down the general Lorentz and flavor structure of the
mean field, and derive the Dirac equation for quarks in that field. The resulting
baryon resonances exhibit an hierarchy of scales: The crude mass is ${\cal O}(N_c)$,
the intrinsic quark excitations are ${\cal O}(1)$, and each intrinsic quark state
entails a finite band of collective excitations that are split as ${\cal O}(1/N_c)$.
We build a (new) theory of those collective excitations, where full dynamics is represented
by only a few constants. In a limiting (but unrealistic) case when the mean field is
spherically-and flavor-symmetric, our classification of resonances reduces to the
$SU(6)$ classification of the old non-relativistic quark model. Although in the real
world $N_c$ is only three, we obtain a good accordance with the observed resonance
spectrum up to 2 GeV.
\end{abstract}



\maketitle

\section{Introduction}

There are currently two main classes of baryon models: three-quark models
with certain interaction between quarks, and models where baryons are treated
as nonlinear solitons of bosonic fields, such that quarks are implicit.
Both points of views have their strong and weak points. The main shortcoming
of the 3-quark models is that they ignore the phenomenologically important admixture
of quark-antiquark pairs ($Q\bar Q$) in a baryon, therefore they are essentially
nonrelativistic. A consistent relativistic picture can be only field-theoretic.

This is the advantage of the solitonic approach that stems from the Skyrme
model~\cite{Skyrme:1961vq,Adkins:1983ya} and includes its more recent incarnation in the
holographic QCD~\cite{Sakai:2004cn,Pomarol:2008aa}. Models of this kind effectively
take care of $Q\bar Q$ pairs in baryons~\cite{Diakonov:2008hc}. However, an obvious
shortcoming is that they do not possess explicit quarks, therefore it is difficult
to address many important issues, for example, what are quark and antiquark (parton)
distributions in nucleons.

In this paper, we suggest an approach that is a bridge between the two. On the one hand,
we operate in terms of quarks and keep contact with the traditional three-quark models.
On the other hand, we do not restrict ourselves to the valence quarks only but allow
for an arbitrary amount of additional $Q\bar Q$ pairs. Our formalism is relativistically
invariant. The key ingredient of our construction is the mean bosonic field for quarks,
which, in fact, is nothing but the ``soliton'' of the second approach.

As in any other solitonic picture of baryons, we formally need to consider the number
of colors $N_c$ as a large algebraic parameter. When $N_c$ is large the $N_c$ quarks
constituting a baryon can be considered in a mean (non-fluctuating) field that does not
change as $N_c\!\to\!\infty$~\cite{Witten:1979kh}. Quantum fluctuations about a mean field
are suppressed as $1/N_c$. While in the real world $N_c$ is only three, we do not expect
qualitative difference in the baryon spectrum from its large-$N_c$ limit. The hope is that
if one develops a clear picture at large $N_c$, and controls at least in principle
$1/N_c$ corrections, its imprint will be visible at $N_c\!=\!3$.

From the large-$N_c$ viewpoint, baryons have been much studied in the past using general
$N_c$ counting rules and group-theoretic arguments, for reviews
see~\cite{Manohar:2001cr,Goity:2005fj,Cohen:2005uv} and references therein. In this framework,
many relations for baryon resonances have been derived, with no reference to the underlying
dynamics. The key new point of this paper is that we suggest a simple underlying physical picture
that results in those relations and disclose the meaning of the otherwise free numerical
coefficients therein. We also derive new relations valid in the large-$N_c$ limit.

This work is in the line with our previous chiral quark--soliton model~\cite{Diakonov:1986yh}
which successfully describes quark and antiquark distributions in
nucleons~\cite{Diakonov:1996sr}, and other properties. In this paper we take off the
previous limitation that the mean field is exclusively the pseudoscalar one, and focus
on baryons resonances rather than on the ground state.

The advantage of the large-$N_c$ approach is that at large $N_c$ baryon physics simplifies
considerably, which enables one to take into full account the important relativistic
and field-theoretic effects that are often ignored. Baryons are not just three
(or $N_c$) quarks but contain additional $Q\bar Q$ pairs, as it is well known
experimentally. The number of antiquarks in baryons is, theoretically, also proportional
to $N_c$~\cite{Diakonov:2008hc}, which means that antiquarks cannot be obtained from
adding one meson to a baryon: one needs ${\cal O}(N_c)$ mesons to explain ${\cal O}(N_c)$
antiquarks, implying in fact a classical mesonic field.

At the microscopic level quarks experience only color interactions, however gluon field
fluctuations are not suppressed if $N_c$ is large; the mean field can be only `colorless'.
An example how originally color interactions are Fierz-transformed into interactions of quarks
with mesonic fields is provided by the instanton liquid model~\cite{Diakonov:1986aj}.
A non-fluctuating confining bag is another example of a `colorless' mean field. A more
modern example of a mean field is given by the 5- or 6-dimensional `gravitational'
and flavor background field in the holographic QCD models.

Since quarks inside baryons are generally relativistic, especially in excited baryons,
we shall assume that quarks in the large-$N_c$ baryon obey the Dirac equation
in a background mesonic field. In fact, the Dirac equation for quarks may be non-local.
All intrinsic quark Dirac levels in the mean field are stable in $N_c$. All negative-energy
levels should be filled in by $N_c$ quarks in the antisymmetric state in color, corresponding
to the zero baryon number state. Filling in the lowest positive-energy level by $N_c$
`valence' quarks makes a baryon. Exciting higher quark levels or making particle-hole
excitations produces baryon resonances. The baryon mass is ${\cal O}(N_c)$, and the
excitation energy is ${\cal O}(1)$. When one excites one quark the change of the mean
field is ${\cal O}(1/N_c)$ that can be neglected to the first approximation.

Moreover, if one replaces one light ($u,d$ or $s$) quark in light baryons by a heavy ($c,b$) one, as in charmed or bottom baryons, the change in
the mean field is also ${\cal O}(1/N_c)$. Therefore, the spectrum of heavy baryons is directly related to that of light baryons.  This fact is
well known for low-lying multiples, see e.g.\cite{Jenkins:1996de}, and recently has been discussed for more general situation in
\cite{Diakonov:2009kw}.

Our approach can be illustrated by the chiral quark--soliton
model~\cite{Diakonov:1986yh,Diakonov:2000pa} or by the chiral bag
model~\cite{Hosaka:1996ee} but actually the arguments of this paper are much more general.
We argue that the mean field in baryons of whatever nature has a definite symmetry,
namely it breaks spontaneously the symmetry under separate $SU(3)_{\rm flavor}$
and $SO(3)_{\rm space}$ rotations but does not change under
simultaneous $SU(2)_{\rm iso+space}$ rotations in ordinary space and a compensating
rotation in isospace~\cite{Diakonov:2008rd,Diakonov:2009kw}.

If the original symmetry (here: flavor and rotational) is spontaneously broken,
it means that the ground state is degenerate: all states obtained by a rotation
have the same energy. In Quantum Mechanics the rotations must be quantized,
which leads to the splitting of the ground state, as well as all one-quark excitations,
by the quantized rotational energy. It implies that each intrinsic quark state,
be it the ground state or a one-quark excitation in the Dirac spectrum, generates
a band of resonances appearing as collective rotational excitations of a given intrinsic
state. The quantum numbers of those resonances, their total number and their splittings
are unequivocally dictated by the symmetry of the mean field. In this paper we present,
for the first time, the theory of the rotational bands stemming from a given
intrinsic one-quark excitation. Assuming the $SU(2)_{\rm iso+space}$ symmetry of
the mean field, we obtain the resonances observed in Nature. Moreover, certain
relations between resonance splittings that are satisfied with high accuracy,
are dynamics-independent but follow solely from the particular symmetry of the mean
field.

In this paper, we do not consider any specific dynamical model but concentrate
mainly on symmetry. A concrete dynamical model would say what is the intrinsic
relativistic quark spectrum in baryons. It may get it approximately correct,
or altogether wrong. Instead of calculating the intrinsic Dirac spectrum of quarks
from a model, we extract it from the experimentally known baryon spectrum
by interpreting baryon resonances as collective excitations about the
state and about the one-quark transitions. However, we show that the needed
intrinsic quark spectrum can be obtained from a natural choice of the mean field
satisfying the $SU(2)_{\rm iso+space}$ symmetry.

In summary, we show that it is possible to obtain a realistic spectrum of baryon
resonances up to 2 GeV, starting from the large-$N_c$ limit.
This means that we are able to find candidates for all rotational $SU(3)$-multiplets generated around intrinsic
quark levels and check large $N_c$ relations between their masses. However, this is not the end
of the story as $SU(3)$-multiplets in nature are splitted due to non-zero mass of the strange
quark. These splittings are not small and not all members of $SU(3)$-mutliplets are known. For this reason
even the contents of lowest $SU(3)$ multiplets is under discussion. We will follow analysis of paper\cite{GP}
(see also \cite{MPS}).

We will assume that mass of the strange quark is small enough and construct  perturbation theory in $m_s$ valid at large
$N_c$. The question of its validity is under discussion as well (see, e.g., \cite{Callan-Klebanov})  but we consider
the success of relations following from this theory (classical Gell-Mann-Okubo or Guadagnini\cite{Guadagnini} ones) as an argument
in favor of this approach. We derive a number of new relations valid at large $N_c$ which are fulfilled with a good accuracy.
Let us note our approach give essentially more relations that were derived in the framework of approach \cite{DJM1} (for mass
relations and other aspects of broken $SU(3)$ symmetry, see \cite{DJM2}). Also we give the dynamical interpretation
of the constants entering mass splitting and provide the formulas which allow one to calculate splittings provided that underlying dynamical model is fixed.

The notion of baryon resonance implies that its width is small. For excited baryons
at large $N_c$ this is not granted --- the width of the baryon is $O(1)$. The width is small compared to
the total mass of baryon ($O(N_c)$) but it appears to be of order the distance between quark
levels. This width is due to transitions between different quark levels with emitting of mesons (e.g.
one pion decay to the ground level) which is not suppressed by $N_c$. We will show below that in the leading order the width
is universal for all rotational band belonging to the given quark level. Non-zero width of the resonance leads also to some
shift in its position. In spite of the fact that this shift is $O(1)$, it is also universal for rotational band and does not ruin rotational
spectra.

Mean field approximation cannot be applied directly to unstable quark levels. The correct definition of
the baryon resonance comes from the consideration of meson-baryon scattering amplitude.
The baryon resonance manifests itself as a pole in the complex plane of the energy with imaginary part being half of the resonance width.
Scattering amplitude at large $N_c$  can be found from meson quadratic form which can be obtained
by integrating out quark degrees of freedom.

We performed this program for exotic pentaquark states in \cite{Diakonov:2008hc} in the framework of Skyrme model but it looks
to be too complicated for the general case of baryons considered in this paper. We will use the fact that only resonances
with relatively small width can be observed and neglect the widths of resonance in order to describe their positions.
Moreover, as we do not consider any dynamical mechanism, positions of quark levels anyway play a role of phenomenological parameters.
At this point our approach is close to the one of the quark model which also neglects influence of the resonance width to its position.

The widths of baryon resonances also have some hierarchy in $N_c$. Decays with transition from one quark level to another are $O(1)$, decays inside the same
rotational band are $O(1/N^2_c)$. In particular, total widths of the baryons belonging to the rotational band of the ground state (like $\Delta$-resonance)
 are only $O(1/N^2_c)$ while total widths of all excited baryons are all $O(1)$. These widths are the same for all excited baryons belonging to the given band up to corrections of $O(1/N_c)$ which can nevertheless be significant at $N_c=3$ (to say nothing about corrections in the mass of the strange quark).

One can try two approaches of adjusting  large $N_c$ limit results to the real $N_c=3$ world. The calculation of physical quantity can be divided
typically in two stages: translating original quantity to some effective rotational operator and calculation of matrix element of effective operator
with wave functions of rotational states  representing given baryon. The first stage requires limit of large $N_c$ in order to avoid the mess of strong interactions.
The second is, in fact, trivial and leads to some $SU(3)$-Clebsch-Gordan coefficients, calculable at any $N_c$. The approach pioneered
in \cite{DJM1}  requires the strict limit $N_c\to\infty$ at both stages. From the other hand in papers \cite{Diakonov:1986yh}
and subsequent ones  we applied another approach: use the limit $N_c\to\infty$ but substitute Clebsch-Gordan coefficients by its value at $N_c=3$. The same logic was, in fact, used also in the original paper of \cite{Adkins:1983ya}. This approach has at least the same accuracy as the first one but allows to avoid large corrections related to the change of Clebsch-Gordan coefficients from $N_c=\infty$ to $N_c=3$. In this paper we will discuss both approaches but use mainly the second one.

The paper is organized as follows. In Section II we discuss the possible symmetry of mean field and come to the conclusion that
it should be a hedgehog one. This is one of the main features distinguishing our model from the quark model (which can be also considered at
$N_c\to\infty$). The quark model assumes that mean field inside the baryon has central symmetry (in majority of the versions it is just the confinement
potential). We believe that this assumption contradicts to the data and the mean meson field (e.g. mean field of the pion) is at least equally
important.

In Section III  we derive Dirac equations in the general hedgehog meson field. One has  to find  intrinsic quark levels in this mean field. To determine
the self-consistent meson field, it is necessary to know also the meson part of Lagrangian. This can be done in the concrete dynamical model. We give the classification
of quark levels  and discuss the possible order of levels in the mean field.

In Section IV we construct the general theory of rotational state around intrinsic quark levels. Previously this theory was discussed for the ground state
baryons; we extend it to arbitrary excited baryon states. We derive formulae for baryon masses and obtain the contents of the $SU(3)$ multiplets entering rotational bands. We also discuss their rotational and quark wave functions.

Section V considers the relation of the $SU(3)$-multiplets  at $N_c\to\infty$ and $SU(6)$ multiplets of the quark model. We explain that there is
one-to-one correspondence between quark model and one-quark excitations in the mean field at $N_c\to\infty$ for negative parity baryons. This
is not true for positive parity: here $SU(6)$ multiplets of the quark model correspond mainly to two-quark excitations. Meanwhile, one-particle excitations still exist and have the same structure as in sector with negative parity. We prefer to use excitations of this type in order to describe experimental data as they should have smaller mass and be narrower than two-particle ones. We leave the quark model picture for parity plus sector and arrive at the description unified for both parities: in order to describe experimental baryon spectra we need 6 levels with grand spin $K=0^\pm, 1^\pm , 2^{\pm}$. We confront this simple picture to the data in Section VI and see that it can accommodate the experimental baryon spectra up to 2 GeV. At the same time it does not predict extra states which are typical for the quark model. Section VI is devoted to the mass splitting inside $SU(3)$-multiplets. This question can be important for identifying original $SU(3)$ multiplets. We concentrate mainly on general relations which are model independent. We formulate our conclusions in Section VII.

We relegate few important questions to the series of Appendices. In Appendix A the simple exactly solvable model is considered. This model was already  investigated in a number of papers; it helps us to illustrate the relations obtained in the main text at $N_c\to\infty$. Appendix B is related to validity of the cranking approximation in the soliton picture; this validity was doubt in some papers. We discuss decays of baryon resonances in Appendix D. The full theory will be published elsewhere. Here we only give $N_c$ counting of the baryon widths due to the different decays and prove the universality of the width in the leading order for the given rotational band.
Appendices C and E are devoted to some technical questions. In particular, in appendix E we give the table of $SU(3)$ Clebsch-Gordan coefficients conforming at $N_c=3$
to the standard conventions.

\section{Symmetry of the mean field}

In the mean field approximation, justified at large $N_c$, one looks for the solutions
of the Dirac equation for single quark states in the background mean field.
In a most general case the background field couples to quarks through all five Fermi variants. If the mean field is stationary in time, it leads to the Dirac eigenvalue equation for the $u,d,s$ quarks in the background field, $H\psi =E\psi$, the Dirac Hamiltonian being schematically
\[
H =\gamma^0\!\biggl(\!-i\partial_i\gamma^i+S({\bf x})+P({\bf x})i\gamma^5
+V_\mu({\bf x})\gamma^\mu+
\]\beq
+A_\mu({\bf x})\gamma^\mu\gamma^5
+T_{\mu\nu}({\bf x})\frac{i}{2}[\gamma^\mu\gamma^\nu]\!\biggr),
\la{DiracH}\eeq
where $S,P,V,A,T$ are the scalar, pseudoscalar, vector, axial, tensor mean fields,
respectively; all are matrices in flavor. In fact, the one-particle Dirac Hamiltonian
\ur{DiracH} is generally nonlocal, however that does not destroy symmetries in which
we are primarily interested. We include the current and the dynamically-generated quarks
masses into the scalar term $S$.

The key issue is the symmetry of the mean field. We assume the chiral limit for
$u,d$ quarks, $m_u=m_d=0$, which is an excellent approximation. We consider
exact $SU(3)$ flavor symmetry as a good starting point. It implies that baryons appear in degenerate $SU(3)$ multiplets
${\bf 8},\;{\bf 10},\ldots$; the splittings inside $SU(3)$ multiplets can be determined
later on as a perturbation in $m_s$, see {\it e.g.} Ref.~\cite{Blotz} and Section VII.

A natural assumption, then, would be that the mean field is flavor-symmetric, and spherically
symmetric. However we know that baryons are strongly coupled to pseudoscalar mesons
($g_{\pi NN}\approx 13$). It means that there is a large pseudoscalar field inside baryons;
at large $N_c$ it is a classical mean field. There is no way of writing down the pseudoscalar
field (it must change sign under inversion of coordinates) that would be compatible with
the $SU(3)_{\rm flav}\times SO(3)_{\rm space}$ symmetry. The minimal extension of spherical
symmetry is to write the ``hedgehog'' {\it Ansatz} ``marrying'' the isotopic and space
axes~\footnote{A.~Hosaka informed us that historically, this {\it Ansatz} for the pion
field in a nucleon appears for the first time in a 1942 paper by Pauli and Dancoff~\cite{PD};
it reappears in 1961 in the seminal papers by Skyrme~\cite{Skyrme:1961vq}.}:
\beq
\pi^a({\bf x})=\left\{\begin{array}{ccc}n^a\,F(r),& n^a=\frac{x^a}{r},& a=1,2,3,\\
0,&&a=4,5,6,7,8.\end{array}\right.
\la{hedgehog}\eeq
This {\it Ansatz} breaks the $SU(3)_{\rm flav}$ symmetry. Moreover, it breaks the symmetry
under independent space $SO(3)_{\rm space}$ and isospin $SU(2)_{\rm iso}$ rotations, and
only a simultaneous rotation in both spaces remains a symmetry, since a rotation in
the isospin space labeled by $a$, can be compensated by the rotation of the space axes.
The {\it Ansatz} \ur{hedgehog} implies a spontaneous (as contrasted to explicit) breaking
of the original $SU(3)_{\rm flav}\times SO(3)_{\rm space}$ symmetry down to the
$SU(2)_{{\rm iso}\!+\!{\rm space}}$ symmetry. It is analogous to the spontaneous breaking
of spherical symmetry by the ellipsoid form of many nuclei; there are many other examples
in physics where the original symmetry is spontaneously broken in the ground state.

We list here all possible structures in the $S,P,V,A,T$ fields, compatible with the
$SU(2)_{{\rm iso}\!+\!{\rm space}}$ symmetry and with the $C,P,T$ quantum numbers of
the fields~\cite{Diakonov:2008rd,Diakonov:2009kw}. The fields below are generalizations of the `hedgehog'
Ansatz \ur{hedgehog} to mesonic fields with other quantum numbers.

Since $SU(3)$ symmetry is broken, all fields can be divided into three categories:\\

\noi\underline{I. Isovector fields acting on $u,d$ quarks}
\vskip -0.25true cm

\bea\la{ud-isovector}
\text{pseudoscalar:}\; P^a({\bf x})\!\!&=&\!\!n^a\,P_0(r),\\
\n
\text{vector, space part:}\; V^a_i({\bf x})\!\!&=&\!\!\epsilon_{aik}\,n_k\,P_1(r),\\
\n
\text{axial, space part:}\; A^a_i({\bf x})\!\!&=&\!\!\delta_{ai}\,P_2(r)+n_an_i\,P_3(r),\\
\n
\text{tensor, space part: }\; T^a_{ij}({\bf x})\!\!&=&\!\!\epsilon_{aij}\,P_4(r)+\epsilon_{bij}\,n_an_b\,P_5(r).
\eea
\noi\underline{II. Isoscalar fields acting on $u,d$ quarks}
\vskip -0.5true cm

\bea\la{ud-isoscalar}
\text{scalar:}\quad S({\bf x})\!\!&=&\!\!Q_0(r),\\
\n
\text{vector, time component:}\quad V_0({\bf x})\!\!&=&\!\!Q_1(r),\\
\n
\text{tensor, mixed components:}\quad T_{0i}({\bf x})\!\!&=&\!\!n_i\,Q_2(r).
\eea

\noi\underline{III. Isoscalar fields acting on $s$ quarks}
\vskip -0.5true cm

\bea\la{s-isoscalar}
\text{scalar:}\quad S({\bf x})\!\!&=&\!\!R_0(r),\\
\n
\text{vector, time components:}\quad V_0({\bf x})\!\!&=&\!\!R_1(r),\\
\n
\text{tensor, mixed components:}\quad T_{0i}({\bf x})\!\!&=&\!\!n_i\,R_2(r).
\eea
All the rest fields and components are zero as they do not satisfy the
$SU(2)_{{\rm iso}\!+\!{\rm space}}$ symmetry and/or the needed discrete $C,P,T$
symmetries. The 12 `profile' functions $P_{0,1,2,3,4,5},\,Q_{0,1,2}$ and $R_{0,1,2}$
should be eventually found self-consistently from the minimization of the mass of
the ground-state baryon. We shall call Eqs. (\ref{ud-isovector}-\ref{s-isoscalar})
the hedgehog {\it Ansatz}. However, even if we do not know those profiles, there are
important consequences of this {\it Ansatz} for the baryon spectrum.

\section{$u,d,s$ quarks in the `hedgehog' field}

Given the $SU(2)_{{\rm iso}+{\rm space}}$ symmetry of the mean field,
the Dirac Hamiltonian for quarks actually splits into two: one for $s$ quarks and the other
for $u,d$ quarks~\cite{Diakonov:2008rd}. It should be stressed that the energy levels for $u,d$
quarks on the one hand and for $s$ quarks on the other are completely different, even in the
chiral limit $m_s\to 0$.

The energy levels for $s$ quarks are classified by half-integer $J^P$ where
$P$ is parity under space inversion, and ${\bf J}={\bf L}+{\bf S}$ is quark angular momentum;
all levels are $(2J+1)$-fold degenerate. The energy levels for $u,d$ quarks are classified by
integer $K^P$ where ${\bf K}={\bf T}+{\bf J}$ is the `grand spin' ($T$ is isospin),
and are $(2K+1)$-fold degenerate.

All energy levels, both positive and negative, are probably discrete owing to confinement.
Indeed, a continuous spectrum would correspond to a situation when quarks are free at large
distances from the center, which contradicts confinement. One can model confinement {\it e.g.}
by forcing the effective quark masses to grow linearly at infinity, $S({\bf x})\to\sigma r$.

The Dirac equation \ur{DiracH} for $s$ quarks in the background field \ur{s-isoscalar}
takes the form of a system of two ordinary differential equations for two functions
$f(r),\;g(r)$ depending only on the distance from the center. The system of equations
depends on the (half-integer) angular momentum of level under considerations, and
on its parity. For $s$-quark levels with parity $P=(-1)^{J-\frac{1}{2}}$,
{\it e.g.} for the levels $J^P=\frac{1}{2}^+,\frac{3}{2}^-,\frac{5}{2}^+,..$,
the system takes the form
\bea\la{s_P=J-}
\left\{\begin{array}{ccc}
E\,f&=&-g'-\frac{J+\frac{3}{2}}{r}\,g+R_0\,f +R_1\,f +R_2\,g  \\
E\,g&=&f'+\frac{-J+\frac{1}{2}}{r}\,f-R_0\,g +R_1\,g +R_2\,f .
\end{array}\right.
\eea
To find an $s$-quark energy level $E$ with these quantum numbers, one has to solve \Eq{s_P=J-}
with the initial condition $f(r)\sim r^{J-\half},\;g(r)\sim r^{J+\half}$, and both functions
decreasing at infinity.

For levels with opposite parity $P=(-1)^{J+\frac{1}{2}}$, {\it e.g.}
$J^P=\frac{1}{2}^-,\frac{3}{2}^+,\frac{5}{2}^-,..$, one has to solve another system:
\bea\la{s_P=J+}
\left\{\begin{array}{ccc}
E\, f&=&- g'-\frac{-J+\frac{1}{2}}{r}\, g+R_0\, f +R_1\, f
+R_2\, g  \\
E\, g&=& f'+\frac{J+\frac{3}{2}}{r}\, f-R_0\, g +R_1\, g
+R_2\, f .
\end{array}\right.
\eea
We note that in the absence of the $R_{1,2}$ fields the energy spectrum is symmetric
under simultaneous change of parity and energy signs.

Dirac equation for $u,d$ quarks in the background fields \urs{ud-isovector}{ud-isoscalar}
is more complicated: one has here a system of four ordinary differential equations. These
equations are direct generalizations of the Dirac equations in the `hedgehog'
field~\cite{Diakonov:1986yh}, and can be derived similarly to how it is done in that reference.

The system of Dirac equations for the radial functions of the states with parity $(-1)^{K+1}$,
namely $K^P=1^+,2^-,...$ has the form
\begin{widetext}
\beq
E\,f =-g'\!-\!\frac{1\!+\!K}{r}\,g\!+\!(Q_0\!+\!Q_1\!+\!P_2\!+\!P_4)f\!+\!(Q_2\!-\!P_1)g
\!-\!\frac{P_0\!-\!P_1}{2K\!+\!1}(g\!+\!b_K\,h)\!+\!\frac{P_3\!+\!P_5}{2K\!+\!1}(f\!+\!b_K\,j),
\la{f1}\eeq
\beq
E\,g=f'\!-\!\frac{K\!-\!1}{r}\,f\!+\!(Q_1\!-\!Q_0\!-\!P_2\!+\!P_4)g\!+\!(Q_2\!-\!P_1) f
\!-\!\frac{P_0\!-\!P_1}{2K\!+\!1}(f\!+\!b_K\,j)\!+\!\frac{P_3\!-\!P_5\!+\!2P_2\!-\!2P_4}{2K\!+\!1}(g\!+\!b_K\,h),
\la{g1}\eeq
\beq
E\,h=j'\!+\!\frac{2\!+\!K}{r}\,j\!+\!(Q_1\!-\!Q_0\!-\!P_2\!+\!P_4)h\!+\!(Q_2\!-\!P_1) j
\!+\!\frac{P_0\!-\!P_1}{2K\!+\!1}(j\!-\!b_K\,f )\!-\!\frac{P_3\!-\!P_5\!+\!2P_2\!-\!2P_4}{2K\!+\!1}(h\!-\!b_K\,g),
\la{h1}\eeq
\beq
E\,j=-h'\!+\!\frac{K}{r}\,h\!+\!(Q_0\!+\!Q_1\!+\!P_2\!+\!P_4)j\!+\!(Q_2\!-\!P_1) h
\!+\!\frac{P_0\!-\!P_1}{2K\!+\!1}(h\!-\!b_K\,g)\!-\!\frac{P_3\!+\!P_5}{2K\!+\!1}(j\!-\!b_K\,f ),
\la{j1}\eeq
\end{widetext}
where $b_K=2\sqrt{K(K+1)}$. The radial functions $f,g,h,j$ refer to partial waves with
$L=K\!-\!1,K,K,K\!+\!1$, respectively, and they behave at the origin as $r^L$.
To find the energy levels for a given $K^P$, one has to solve
these equations twice: once with the initial condition
$f(r_{\rm min})\sim r_{\rm min}^{K\!-\!1}$, all the rest functions being put to zero
at the origin, and another time with the initial
condition $h(r_{\rm min})\sim r_{\rm min}^K$, with all the rest functions zeroes, $r_{\rm min}\to 0$.
Evolving the functions according to the equations numerically up to some asymptotically
large $r_{\rm max}$ one finds two sets of functions $(f_1,g_1,h_1,j_1)$ and
$(f_2,g_2,h_2,j_2)$. The energy levels are found from the zeroes of two (equal)
determinants $f_1h_2-f_2h_1=g_1j_2-g_2j_1$.

For states with parity $(-1)^{K}$, namely $K^P=1^-,2^+,..$ the system of Dirac
equations is:
\begin{widetext}
\beq
E\,f=-g'\!-\!\frac{1+K}{r}\,g\!+\!(\!Q_1\!-\!Q_0\!+\!P_2\!-\!P_4)f\!-\!(Q_2\!+\!P_1)g
\!+\!\frac{P_0\!+\!P_1}{2K\!+\!1}(g\!+\!b_K\,h)\!+\!\frac{P_3\!-\!P_5}{2K\!+\!1}(f\!+\!b_K\,j),
\la{f2}\eeq
\beq
E\,g=f'\!-\!\frac{K\!-\!1}{r}\,f\!+\!(Q_0\!+\!Q_1\!-\!P_2\!-\!P_4)g\!-\!(Q_2\!+\!P_1)f
\!+\!\frac{P_0\!+\!P_1}{2K\!+\!1}(f\!+\!b_K\,j)\!+\!\frac{P_3\!+\!P_5+\!2P_2\!+\!2P_4}{2K\!+\!1}(g\!+\!b_K\,h),
\la{g2}\eeq
\beq
E\,h=j'\!+\!\frac{2\!+\!K}{r}\,j\!+\!(Q_0\!+\!Q_1\!-\!P_2\!-\!P_4)h\!-\!(Q_2\!+\!P_1) j
\!-\!\frac{P_0\!+\!P_1}{2K\!+\!1}(j\!-\!b_K\,f )\!-\!\frac{P_3\!+\!P_5+\!2P_2\!+\!2P_4}{2K\!+\!1}(h\!-\!b_K\,g),
\la{h2}\eeq
\beq
E\,j=-h'\!+\!\frac{K}{r}\,h\!+\!(\!Q_1\!-\!Q_0\!+\!P_2\!-\!P_4)j\!-\!(Q_2\!+\!P_1) h
\!-\!\frac{P_0\!+\!P_1}{2K\!+\!1}(h\!-\!b_K\,g)\!-\!\frac{P_3\!-\!P_5}{2K\!+\!1}(j\!-\!b_K\,f),
\la{j2}\eeq
\end{widetext}
where again $f\sim r^{K\!-\!1},g\sim r^K,h\sim r^K, j\sim r^{K\!+\!1}$, and the levels
are found by the same trick. The fields $Q_{1,2}$ and $P_{0,2,3}$ break symmetry with
respect to simultaneous change of parity and energy signs.
\begin{widetext}
\begin{center}
\begin{figure}[htb]
\begin{minipage}[t]{0.42\textwidth} 
\includegraphics[width=\textwidth]{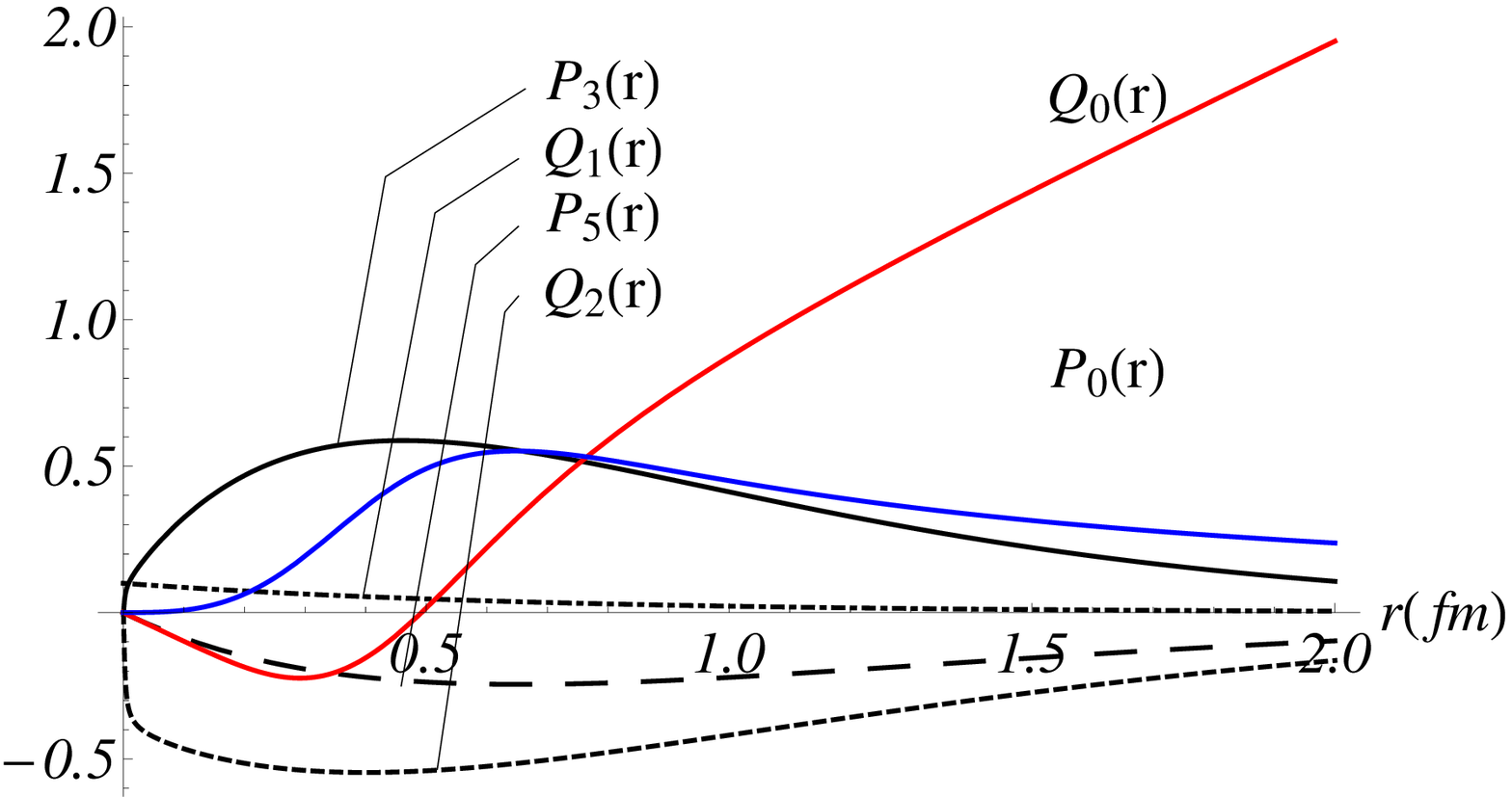}
\end{minipage}
\hspace{1.5cm}
\begin{minipage}[t]{0.42\textwidth} 
\includegraphics[width=\textwidth]{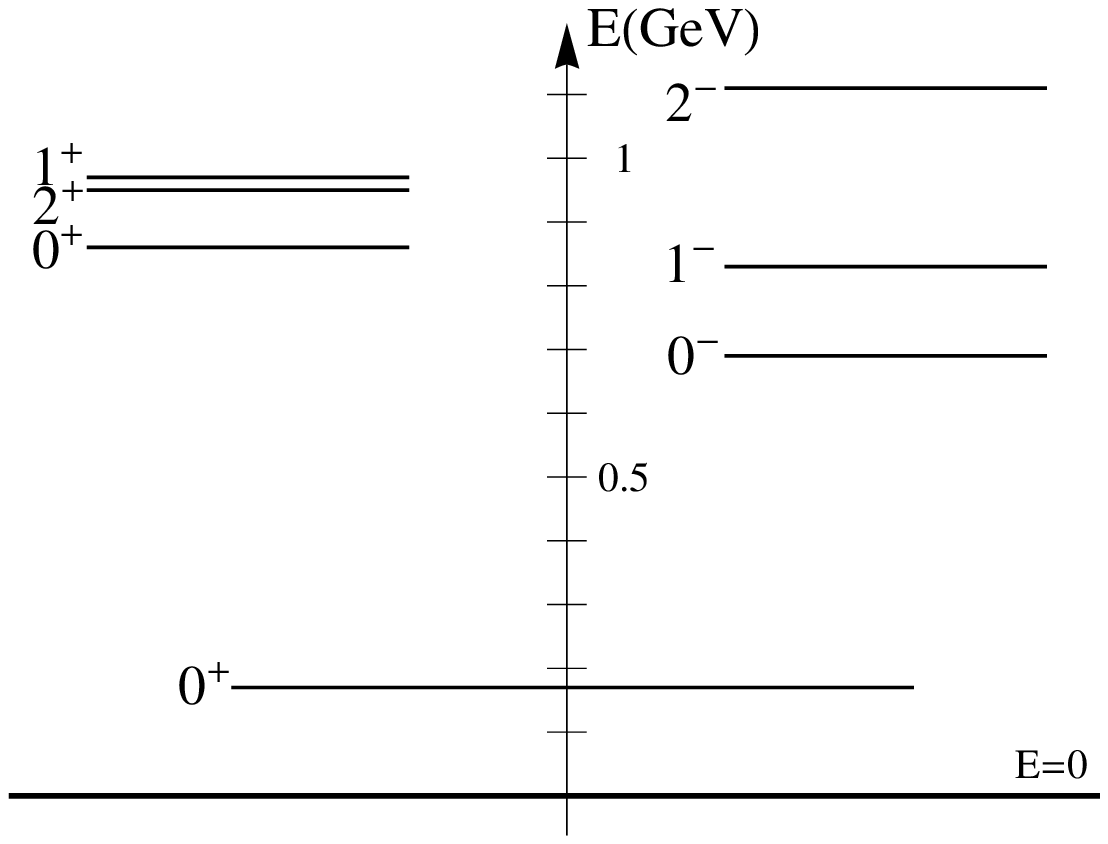}
\end{minipage}
\caption{(Color online) An illustrative example of intrinsic quark levels with quantum numbers $K^P$ (right)
generated by the mean fields shown in the left panel.}
\label{fig:1}
\end{figure}
\end{center}
\end{widetext}
The case $K=0$ is special, since the angular momentum is restricted to only one value $J=K+\frac{1}{2}=\frac{1}{2}$. It means that $g=f=0$, and
the system of eqs.(\ref{f1})-(\ref{j1}) for the $K^P=0^-$ level reduces to two equations:
\begin{widetext}
\bea\nn
E\,j&=&-h'+(Q_0+Q_1+P_2-P_3+P_4-P_5)j+(P_0-2P_1+Q_2)h, \\
E\,h&=&j'+\frac{2}{r}\,j+(-Q_0+Q_1-3P_2-P_3+3P_4+P_5)h+(P_0-2P_1+Q_2)j
\la{App_Sys2(0-)}\eea
with $h\sim r^0,j\sim r^1$.
Similarly, to find the $K^P=0^+$ levels one has to solve only two equations:
\bea\nn
E\,j&=&-h'+(-Q_0+Q_1+P_2-P_3-P_4+P_5)j-(P_0+2P_1+Q_2)h, \\
E\,h&=&j'+\frac{2}{r}\,j+(Q_0+Q_1-3P_2-P_3-3P_4-P_5)h-(P_0+2P_1+Q_2)j.
\la{App_Sys2(0+)}\eea
\end{widetext}

In Fig.~1 we show an example of quark levels obtained from a `natural' choice of external fields $Q_{0\!-\!2},P_{0\!-\!5}$. We take a confining
scalar field $S(r)=\sigma r$ with a standard string tension $\sigma = (0.44\,{\rm GeV})^2$, and a topological chiral angle field
$P(r)=2\arctan(r_0^2/r^2)$ such that the profile functions introduced in \Eqs{ud-isovector}{ud-isoscalar} are $Q_0(r)=S(r)\cos
P(r),\;P_0(r)=S(r)\sin P(r)$; the other profile functions are exponentially decaying at large distances. The external fields are shown in
Fig.~1, left, and the resulting quark levels with various $K^P$ are shown in Fig.~1, right. These or similar levels dictate the masses of baryon
resonances.

According to the Dirac theory, all {\em negative}-energy levels, both for $s$ and $u,d$ quarks,
have to be fully occupied, corresponding to the vacuum. It means that there must be exactly
$N_c$ quarks antisymmetric in color occupying all degenerate levels with $J_3$ from $-J$
to $J$, or $K_3$ from $-K$ to $K$; they form closed shells. Filling in the lowest level with
$E>0$ by $N_c$ quarks makes a ground state baryon, see Fig.~2. A similar picture arises
in the chiral bag model~\cite{Hosaka:1996ee}. Excited baryons can be related to different 1,2,3-quark excitations
to the other levels. We will try to advocate the point of view that known baryon resonances below 2 GeV are related
to one quark excitations only.

\begin{figure}[htb]
\includegraphics[width=0.42\textwidth]{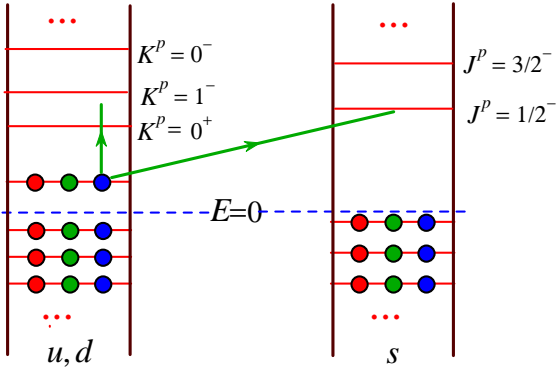}
\caption{(Color online) Filling $u,d$ and $s$ shells for the ground-state baryon. Excitations of one valence quarks
describe baryon resonances.}
\label{fig:2}
\end{figure}

The mass of a baryon is the aggregate energy of all filled states, and being a functional
of the mesonic field, it is proportional to $N_c$ since all quark levels are degenerate
in color. Therefore quantum fluctuations of mesonic field in baryons are suppressed
as $1/N_c$ so that the mean field is indeed justified.

\section{Rotational bands about intrinsic quark levels}
\la{Theory}

Every intrinsic level is accompanied by the rotational band of the states.
It  appears as a result of the quantization of the slow rotations both in the flavor and
ordinary space. The theory of rotational bands over the ground state
was developed years ago\cite{Diakonov:1986yh} but for excited states and for general case of the
mean field it has some specifics.

The original symmetry of the theory in the chiral limit is $SU(3)_{\rm flav}\times SO(3)_{\rm space}$. Both symmetries are broken by the
`hedgehog' {\it Ansatz} of the mean field, so the soliton transforms under the space and flavor rotations non-trivially. However, the energy of
the rotated soliton is the same as the original one. For this reason constant rotations are zero modes and should be taken into account exactly.

\subsection{Ground states}

Rotations slowly depending on time split the energy level into the rotational band. It is convenient to describe this effect with an effective
Lagrangian depending on collective coordinates which are rotational matrices.

Let $R(t)$ be an $SU(3)$ matrix describing  slow rotation in flavor space, and ${\cal S}(t)$ be an
$SU(2)$ matrix for slow space (and spin) rotation. They rotate quark wave functions $\phi^{\alpha i}(\bm x)$
($\alpha=1\ldots 3$ is flavor, $i$=1\ldots 2 -- spin indices) in the given mean field as:
\[
\tilde\phi^{\alpha i}_n(x)= R^\alpha_{\alpha'}(t){\cal S}^i_{i'}(t)\phi^{\alpha'i'}_n(O(t)\bm{x}),
\]\beq
 O_{ik}(t)=\half{\rm Tr}\left[{\cal S}^+(t)\sigma_k{\cal S}(t)\sigma_i\right]
\la{rot-of-ferm}
\eeq
Then it is easy to see that simultaneous transformation of the meson fields
\[
\tilde{P}^a(\bm x)=O_{ab}[R]P^b(O({\cal S}){\bm x})
\]\[
\tilde{V}^{ai}(\bm x)=O_{ab}[R]O_{ij}[{\cal S}]V^{bj}(O({\cal S}){\bm x}),
\]\beq
\tilde{A}^{ai}(\bm x)=O_{ab}[R]O_{ij}[{\cal S}]A^{bj}(O({\cal S}){\bm x}),
\la{rot-of-mes}
\eeq
(and so on) leaves invariant Dirac equation in the mean field provided that matrices $R$ and ${\cal S}$ are constant in time.

Let us integrate out quarks. Effective action of the theory is a sum of meson Lagrangian and the contribution of constituent quarks which is the
determinant of the Dirac equation in the mean field: \beq S_{\text{eff}}= \int\! dt\; {\cal L}(M)-i \sum_c \Sp_{\text{occup}}{\rm
Log}\left\{i\frac{\partial}{\partial t}-{\cal H}[M]\right\} \la{SeffM} \eeq Here sum implies the summation in color indices and
${\text{Sp}\{\ldots\}}$  is running over all {\em occupied} states. Since meson field $M$ and the Hamiltonian ${\cal H}$ are
color blind, the sum in color produces the factor $N_c$ for ground state. For the 1-particle excitations one term in this
sum corresponds to some different filling of the levels.

Slow rotations ${\cal S}(t),R(t)$ are the particular cases of the quantum fluctuations of the general meson field $M$. Usually quantum
fluctuations are suppressed in the limit of large $N_c$. Rotations are not suppressed as they are zero modes. Only their frequencies are small
in $N_c$.

Let us parametrize the general meson field as $M=\bar M+\delta M$ (where $\bar M(\bm x)$ is a time independent mean field and $\delta M(\bm x,t)$ are quantum fluctuations) and calculate the effective action \eq{SeffM} on the set of slowly rotated states \urs{rot-of-ferm}{rot-of-mes}
\footnote{We imply here that some conditions are imposed on quantum fluctuations $\delta M$ in order to make them orthogonal to rotations ${\cal R}(t)$, ${\cal S}(t)$. We do not need to specify this procedure. Let us mention, however, that quadratic form $W$ and mixed form ${\cal K}$ in \eq{quadr-gen} are restricted to this subspace of $\delta M$.
}:
\[
S_{\text{eff}}= \int\! dt\; {\cal L}_{\text{meson}}(\bar M+\delta M,\tOmega,\tomega)-
\]\beq
-i \sum_c \Sp_{\text{occup}}{\rm  Log}\left\{i\frac{\partial}{\partial t}-{\cal H}[\bar M+\delta M] -\tOmega_a t_a -\tomega_i j_i \right\}
\la{SeffRot} \eeq Here $\tOmega_a$ and $\tomega_i$ are flavor and angular frequencies in the body-fixed frame: \beq \tOmega_a = -i
\Tr\left[R^+\dot{R}\lambda_a\right], \qquad \tomega_i = -i\Tr\left[{\cal S}^+ \dot{{\cal S}}\sigma_i\right] \eeq ($\lambda_a$ are Gell-Mann
flavor matrices and $\sigma_i$ are Pauli spin matrices), $t_a$ and $j_i$ are one-particle operators of flavor and total angular momenta:
\beq t_a = \half \lambda_a, \qquad j_i=s_i+l_i=\half\sigma_i +i\varepsilon^{ikl}x_k\frac{\partial}{\partial x_l} \eeq

Next we expand \eq{SeffRot} in small $\delta M$, $\tOmega, \tomega$. The linear term should be absent, as mean field  $\bar
M(\bm x)$ is a solution of equations of motion. There is a famous exclusion from this rule --- Witten-Wess-Zumino term which is linear in
$\Omega_8$ and proportional to the baryon charge $B$ of the state: \beq \delta S^{(1)} = -\frac{N_c}{2\sqrt{3}}\int\! dt \tOmega_8 \eeq The
second order correction is in general:
\[
\delta S^{(2)}\!=\!\half \int\!\! d^4x\; \delta M W \delta M + \int\!\! d^4x \left(\delta M \K^a_\Omega \tOmega_a+
\delta M \K^i_\omega\tomega_i\right)\!-
\]\beq
-\half\int\!dt\left[ I^{(\Omega\Omega)}_{ab} \tOmega_a\tOmega_b+ I^{(\omega\omega)}_{ab} \tomega_i\tomega_j +I^{(\omega\Omega)}_{ai}
\tOmega_a\tomega_i\right] \la{2order} \la{quadr-gen} \eeq Here first term is a quadratic form for the quantum fluctuations which are not
rotations, second term describes the mixing of rotations and other quantum fluctuations,  third  term is a generic quadratic form for space and
flavor rotations. All coefficients in \eq{quadr-gen} ($W,\K,I$) are proportional to $N_c$. Thus the quantum fluctuations are $\delta
M=O(1/\sqrt{N_c})$. As to the frequencies $\tOmega,\tomega$ we will see that they are $\tOmega,\tomega=O(1/N_c)$

We are interested in the collective rotational Lagrangian, i.e. the Lagrangian depending only on angular and flavor frequencies. There are two sources for such a Lagrangian. First, the Lagrangian comes from the immediate expansion of the
original action \eq{SeffRot}. Second, in presence of mixing the rotation Lagrangian can arise as a result
of integration over non-rotational quantum fluctuations of the meson field $\delta M$. Indeed, in
the second case the correction to the mean field:
\beq \delta M = W^{-1}\left[ \K^a_\Omega \tOmega_a + \K^i_\omega \tomega_i\right]
\la{deltaM}
\eeq
is  of the first order in frequencies and should be accounted in the leading order rotational Lagrangian:
\[
S^{(2)}_{\text{rot}} =\! -\!\int\!dt\left[
\half \tOmega_a{\cal I}^{(\Omega\Omega)}_{ab}\tOmega_b
+\half  \tomega_i{\cal I}^{(\omega\omega)}_{ij}\tomega_j +\half  \tOmega_a
{\cal I}^{(\omega\Omega)}_{ai}\tomega_i\right]
\]\[
{\cal I}^{(\Omega\Omega)}_{ab}= {I}^{(\Omega\Omega)}_{ab}+\K^a_\Omega W^{-1}\K^b_\Omega,
\]\[
{\cal I}^{(\omega\omega)}_{ij}={I}^{(\omega\omega)}_{ij}+\K^i_\omega W^{-1}\K^j_\omega,
\]\beq
{\cal I}^{(\omega\Omega)}_{ai}=I^{(\omega\Omega)}_{ai}+\K^i_\omega W^{-1}\K^a_\Omega+ \K^a_\Omega W^{-1}\K^i_\omega
\eeq
i.e. the mixing leads to the renormalization of the moments of inertia. It is essential that the terms arising from mixing are of the same order in $N_c$ ( as
$\K\sim O(N_c)$ and $W\sim  O(N_c)$) and contribute to the collective action.

This phenomenon is well-known from the nuclear physics. The approximation with the mixing is neglected is called the
{\em cranking} one \cite{Inglis}. The importance of the mixing was pointed out by Thouless-Valatin \cite{T-V}. The mixing of the rotations and
quantum fluctuations, however, is absent in many relativistic theories (at least this is true for models based only on pions (see Appendix B).
In such theories cranking approximation is exact.

Cranked moments of inertia $I^{(\Omega\Omega)}_{ab}$, $I^{(\omega\omega)}_{ij}$, $I^{(\omega\omega)}_{ai}$   consist of two
parts, fermion and meson ones. To obtain the meson part we substitute rotated meson fields \eq{rot-of-mes}
in the mean field approximation to the meson Lagrangian. If the last Lagrangian contains some time derivative, we will get some terms
quadratic in frequencies $\tOmega$, $\tomega$ (one should neglect higher terms) which are the contributions to the moments of inertia.

The quark part of moments of inertia can be obtained expanding fermion determinant of \eq{SeffRot} in $\tOmega$, $\tomega$. Corresponding part
of the moments of inertia is given by well-known Inglis expression \cite{Inglis}:
\beq
\left(I^{(\Omega\Omega)}_{ab}\right)_{\text{q}}=
\frac{N_c}{2}\sum_{n,m}\frac{\langle n|t_a|m\rangle \langle
m|t_b|n\rangle+\langle n|t_b|m\rangle \langle m|t_a|n\rangle}{\varepsilon_m-\varepsilon_n}
\la{Inglis-inertia}
\eeq
($|n\rangle$ are occupied
1-quark states, $|m\rangle$ are non-occupied states) and analogous expressions for ${I}^{(\omega\omega)}_{ij}$ and ${I}^{(\omega\Omega)}_{ij}$
with flavor generator $t_a$ replaced by total quark angular momentum $j_i$.

Hedgehog symmetry of the mean field leads to the following relations between different moments of inertia:
\[
{\cal I}^{(\Omega\Omega)}_{ab} =\left\{
\begin{array}{cc}
I_1 \delta_{ab}, & a,b=1\ldots 3 \\
I_2 \delta_{ab}, & a,b=4\ldots 7 \\
0, & a,b=8
\end{array}
\right.,
\]\beq
{\cal I}^{(\omega\Omega)}_{ai}=-2 I_1\delta_{ai} \qquad {\cal I}^{(\omega\omega)}_{ij}=I_1\delta_{ij}
\eeq
 and hence the quadratic part of the
rotational action reduces to: \beq S^{(2)}_{rot}=-\int\! dt\sum_{i=1}^3 \frac{I_1}{2}(\tOmega_i-\tomega_i)^2 + \sum_{a=4}^7
\frac{I_2}{2}\tOmega_a^2 \eeq This fact does not depend on the origin of the rotational Lagrangian. In particular this result can be
checked for the quark part \eq{Inglis-inertia} (see, e.g. \cite{Diakonov:1986yh}).

The complete rotational Lagrangian:
\beq
{\cal L}_{rot}= \sum_{i=1}^3 \frac{I_1}{2}(\tOmega_i-\tomega_i)^2 + \sum_{a=4}^7 \frac{I_2}{2}\tOmega_a^2 +
\frac{BN_c}{2\sqrt{3}}\tOmega_8
\la{L-ground}
\eeq
is a Lagrangian for some specific spherical top both in the flavor and usual space. We calculate operators of angular $\tilde{\bm J}$ and flavor momenta $\tilde{\bm T}$:
\[
\tilde{\bm J} =-\half\Tr\left[{\cal S}\bm\sigma\frac{\delta}{\delta{\cal S}}
\right]= \frac{\partial {\cal L}_{rot}}{\partial \bm\omega} = I_1 (\bm\omega-\bm \Omega)
\]
\[
\tilde{T}_a =-\half\Tr\left[R\lambda_a\frac{\delta}{\delta R}
\right]= \frac{\partial {\cal L}_{rot}}{\partial \Omega_a} =
\]\beq
=\left\{
\begin{array}{ll}
I_1 (\Omega_a-\omega_a), &  a=1\ldots 3 \\
I_2 \Omega_a, & a=4\ldots 7 \\
\frac{N_c}{2\sqrt{3}}, & a=8 \\
\end{array}
\right. \la{tilde-momenta} \eeq We see that the following quantization rules applied to the rotational bands of ground state baryons: \beq
\tilde{\bm J}+\tilde{\bm T}=0, \qquad \tilde{T}_8=\frac{N_c}{2\sqrt{3}} \la{quant-ground} \eeq The second is celebrated Witten quantization rule
\cite{Adkins:1983ya}  which claims that hypercharge in the body-fixed frame is $\tilde Y=\frac{2}{\sqrt{3}} \tilde{T}_8=N_c/3$. It is completely
the result of the hedgehog symmetry and fact that $N_c$ valence quarks with the hypercharge $\tilde Y=1/3$ are put to some
bound state in the sector of $u,d$-quarks.

The Hamiltonian of rotations determined from \eq{L-ground} should be expressed in terms of momenta $\tilde T$,
$\tilde J$:
\[
{\cal H}_{rot}=\sum_{a=1}^{3} \frac{{\tilde T}_a^2}{2I_1} +\sum_{a=4}^{7} \frac{{\tilde T^2}_a}{2I_2}=
\]\beq
= \frac{c_2(r)-\tilde{T}(\tilde{T}+1)-\frac{3}{4}\tilde{Y}}{2I_2}+\frac{\tilde{T}(\tilde{T}+1)}{2I_1}
 \la{en-ground} \eeq
 Here
$c_2(r)=\sum_a{\tilde T}^2_a$ is Casimir operator in the  $SU(3)$ representation $r$. It is easy to determine also the
collective wave function which is an eigenfunction of the Hamiltonian and operators of momenta in the lab fixed frame: \beq T_a
=-\half\Tr\left[\lambda_aR\frac{\delta}{\delta R}\right], \qquad \bm J =-\half\Tr\left[\bm\sigma{\cal S}\frac{\delta}{\delta{\cal S}} \right]
\eeq Wave function is a product of two Wigner ${\cal D}$-functions, one for $SU(3)$ and one for $SU(2)$ group:
\[
\Psi_{rot}(R,{\cal S}) =\sqrt{{\rm dim}(r)(2J+1)}\times
\]\[
\times\sum_{\tilde{T},\tilde{T}_3,\tilde{J}_3} C^{00}_{\tilde{T}\tilde{T}_3\;J\tilde{J}_3} {\cal
D}^{(r)}_{\tilde{Y}\tilde{T}\tilde{T}_3;YTT_3}(R^+){\cal D}^{J}_{\tilde{J}_3;J_3}({\cal S}^+)=
\]\beq
=\sqrt{{\rm dim}(r)}(-1)^{J+J_3}{\cal D}^{(r)}_{\tilde{Y}J,-J_3;Y T T_3}({\cal S}R^+) \la{rotwf-ground} \eeq This function is an eigenfunction
of  spin $\bm J^2=\tilde{\bm J}^2=\tilde{\bm T}^2$, $J_3$, isospin $\bm T^2$ and $T_3$ and hypercharge $Y$, index $(r)$ labels the $SU(3)$
representations with dimension dim($r$). According to \eq{quant-ground} the hypercharge $\tilde Y=N_c/3$. At last Clebsch-Gordan
coefficient  $C^{00}_{\tilde{T}\tilde{T}_3\;J\tilde{J}_3}$ sums the isospin $\tilde{\bm T}$ and the angular momentum
$\tilde{\bm J}$ to zero in order to obey  quantization rule \eq{quant-ground}. In fact,  rotational  wave
function depends only on the combination $R{\cal{S}}^+$. This is natural because owing to the hedgehog symmetry flavor isospin rotation can
be compensated by space one.

\subsection{1-quark excited states}

Let us proceed now with 1-quark excitations, i.e. excitations where only one quark  is taken from the ground level and
is put to some excited level.  The effective Lagrangian \eq{SeffRot} is only slightly changed: one term in the sum over
 $N_c$ quarks has a different scheme of occupied levels. The other $N_c-1$ terms, however, remain the same. This means that in
the leading order in $N_c$ the mean field does not change (the correction to the mean field is $O(1/N_c)$). This is also true for moments of
inertia $I_1$ and $I_2$
--- they acquire corrections $O(1)$ as compared to the leading order $O(N_c)$. Therefore  effective rotational Lagrangian \eq{L-ground} is valid
also in this case. However, additional {\em linear} terms in frequencies $\Omega$ and $\omega$ can appear. The reason is that mean field is a solution of
equations of motion only for a ground state and not for excited states. Hence, there is no reason why linear
terms in a perturbation (which is a rotation in this case) should  be absent. The corresponding linear terms are of the form:
\beq \delta{\cal
L}_{\text{rot}}=\langle\exc|(\bm\omega\cdot\bm j + \bm\Omega\cdot\bm t)+\delta M |\exc\rangle \la{L-lin}
\eeq
where  last term accounts for possible change of the contribution of the correction \eq{deltaM} to the mean field as due to rotations. This correction should be also
calculated only in the ground state (it is determined mainly by rotation of other $N_c-1$ quarks) and assumed to be already known. It is also
linear in frequencies $\tomega, \tOmega$.

Excited states are usually degenerate. Indeed, excitations in the $s$-quark sector have degeneracy $2S+1$ (where $S=\half,\frac{3}{2},\ldots$ is a
total momentum of the state), excitations in the sector of $u,d$-quarks are degenerate $2K+1$ fold where $K$ is the grand spin of
the state. Any of degenerate states or their mixture can be taken as an excitation. We define:
\beq |\exc\rangle = \sum \chi_{K_3}
|KK_3JL\rangle \la{exclev}
\eeq
(we are dealing now with $K\neq 0$ excitations for definiteness). Here $|KK_3JL\rangle$ is the wave function of some excited state with grand spin $K$ and
projection $K_3$,  $\chi_{K_3}$ are amplitudes for different values of projection. Energy does not depend on $\chi_{K_3}$. Hence it is a new zero mode and should
be considered as a collective coordinate together with $S$ and $R$. Effective rotational Lagrangian should be written for $\chi_{K_3}$
slowly changing with time, evidently the complete Lagrangian is
\beq {\cal L}_{\exc}[\chi,R,S]= \sum_{K_3} \chi^+_{K_3}i\frac{\partial}{\partial
t}\chi_{K_3} + {\cal L}_{\text{rot}}+\delta{\cal L}_{\text{rot}} \la{L-excited} \eeq
where ${\cal L}_{\text{rot}}$ is the rotational Lagrangian for the ground state, \eq{L-ground}.

Plugging \eq{exclev} into \eq{L-lin} we obtain:
\[
\delta{\cal L}_{\text{rot}} = \sum_{K_3K'_3}\! \chi^+_{K'_3}\chi_{K_3}\biggr[\langle KK_3JL|(\bm\omega\cdot\bm j+\bm\Omega\cdot\bm t)|  KK_3'JL\rangle
\!+
\]\beq
\left.
+(\bm\omega-\bm\Omega)\langle KK_3'JL|\frac{\partial \delta M}{\partial\bm\omega}|KK_3JL\rangle
\right]
\eeq
We used here that  the fluctuation of the meson field should depend only on the difference
of flavor and space frequencies due to the hedgehog symmetry of the ground state: $\delta M\sim \bm\omega-\bm\Omega$.

One-quark flavor momentum $\bm t$ and angular momentum $\bm j$ do not conserve in the hedgehog field. Nevertheless,
since they transformed as vectors under simultaneous flavor and spin rotations their matrix elements should be proportional to the matrix elements of the conserved quantity --- grand spin $\bm K$:
\[
\langle KK_3jl|\bm t |  KK_3'jl\rangle =a_K\langle KK_3jl|\bm K |  KK_3'jl\rangle,
\]\[
\langle KK_3jl|\bm j |  KK_3'jl\rangle =(1-a_K) \langle KK_3jl|\bm K |  KK_3'jl\rangle
\]\beq
\langle KK_3jl|\frac{\partial \delta M}{\partial\bm\omega}|  KK_3'jl\rangle = \zeta_K \langle KK_3jl|\bm K |  KK_3'jl\rangle
\la{mat-K}
\eeq
(the second of these relations follows from $\bm j+\bm t=\bm K$) where $a_K$  and $\zeta_K$
are some constants specific for given excited level. Using explicit expressions for wave functions of the levels with
given $K$ one can derive the following relation for $a_K$:
\[
a_K= \frac{K+1-c_K(2K+1)}{2K(K+1)},
\]\beq
 c_K= \frac{\int\! dr\; r^2\left(h^2(r)+j^2(r) \right)}{\int\! dr\; r^2\left(h^2(r)+j^2(r)+g^2(r)+f^2(r) \right)}
\la{toapp}
\eeq
where $h,j,f,g$ are radial wave functions --- solutions of the Dirac equation (\ref{f1}-\ref{j1}).
 Coefficient $c_K$  ($0<c_K<1$) is measuring the admixture of the state $j=K+\half$ in the wave function of the level with given $K$ (complete
 wave function consists of two states $j=K\pm\half$) (see Appendix C).

In general case it is not possible to calculate the coefficient $\zeta_K$.  In
particular, it depends on the form of the meson Lagrangian. The coefficient $\zeta_K$ renormalizes the value of  $a_K$. Fortunately, the correction to the mean
field $\delta M$ is zero in the wide class of theories.

Collecting all terms we obtain the collective Lagrangian for 1-quark excitations in sector of $u,d$-quarks:
\[
{\cal L}_{K}[\chi,R,S]= \sum_{K_3} \chi^+_{K_3}i\frac{\partial\chi_{K_3}}{\partial t}+
\frac{N_c}{2\sqrt{3}}\tOmega_8+
\]\[
+[(1-\tilde{a}_K)\bm\omega+\tilde{a}_K\bm\Omega] \sum_{K_3 K'_3}\chi^+_{K_3}\chi_{K'_3} \langle KK_3jl|\bm K |  KK_3'jl\rangle+
\]\beq
+\sum_{i=1}^3 \frac{I_1}{2}(\tOmega_i-\tomega_i)^2 + \sum_{a=4}^7 \frac{I_2}{2}\tOmega_a^2, \qquad \tilde{a}_K=a_K-\zeta \la{L-exc-K}
\eeq

Quantization of $\chi_{K_3}$ with Lagrangian \eq{L-exc-K} is trivial. Due to the presence of collective variable $\chi_{K_3}$ the quantity:
\beq
\sum_{K_3 K'_3}\chi^+_{K_3}\chi_{K'_3} \langle KK_3jl|\bm K |  KK_3'jl\rangle=\hat{\bm K}
\eeq
behaves as quantum operator of the angular momentum $K$. Differentiating in $\omega,\Omega$ we obtain the momenta in the body fixed frame:
\[
\tilde{\bm J}= I_1(\bm\omega-\bm\Omega)+(1-\tilde{a}_K)\hat{\bm K}, \qquad
\tilde{\bm T}= I_1(\bm\Omega-\bm\omega)+\tilde{a}_K\hat{\bm K}
\]\beq
\tilde{T}_a=I_2 \tilde \Omega_a,\; (a=4\ldots 8), \qquad \tilde{T}_8 = \frac{N_c}{2\sqrt{3}}
\la{forms}
\eeq
It leads to the following quantization conditions instead of \eq{quant-ground}:
\beq
\tilde{\bm T}+\tilde{\bm J}=\hat{\bm K}, \qquad \tilde{Y}=\frac{N_c}{3}
\la{quant-K}
\eeq
Constructing now the Hamiltonian from the Lagrangian \eq{L-exc-K} we obtain:
\[
{\cal H}_K= \frac{1}{2I_2}\sum_{a=4}^{7} (\tilde{T}_a )^2 +\frac{(\tilde{\bm T}-\tilde{a}_K\hat{\bm K})^2}{2I_1}=
\]\beq
=\frac{1}{2I_2}\sum_{a=4}^{7} (\tilde{T}_a )^2 +\frac{(\tilde{\bm T}-\tilde{a}_K(\tilde{\bm J}+\tilde{\bm T}))^2}{2I_1} \eeq
Energy levels are:
\[
{\cal E}_K= \frac{c_2(r)-\tilde{T}(\tilde{T}+1)-\frac{3}{4}\tilde{Y}^2}{2I_2}+\frac{1}{2I_1}\left[\tilde{a}_K J(J+1)
+
\right.
\]\beq
\left.
+ (1-\tilde{a}_K)\tilde{T}(\tilde{T}+1)
-\tilde{a}_K(1-\tilde{a}_K)K(K+1)\right]
\la{en-K}
\eeq
We used here that $\tilde{J}=J$. Available spins are determined by the quantization rule \eq{quant-K}:
$J=|\tilde{T}-K|\ldots \tilde{T}+K$.

It is easy to construct the collective wave function. For this case it depends on ${\cal S}$, $R$ and $\chi_{K_3}$:
\[
\Psi_{K}(R,{\cal S},\chi) = \sqrt{\frac{{\rm dim}(r)(2J+1)}{2K+1}}\times
\]\beq
\times\sum_{\tilde{T},\tilde{T}_3,\tilde{J}_3} C^{KK_3}_{\tilde{T}\tilde{T}_3\;J\tilde{J}_3} {\cal
D}^{(r)}_{\tilde{Y}\tilde{T}\tilde{T}_3;YTT_3}(R^+){\cal D}^{J}_{\tilde{J}_3;J_3}({\cal S}^+)\chi_{K_3}
 \la{rotwf-K}
 \eeq
 This wave function is
an eigenfunction of hypercharge $Y$, isospin $T$ and its projection $T_3$ as well as spin $J$ and its projection $J_3$. In fact, it is
completely fixed by the symmetry and quantization requirements \eq{quant-K}.

\subsection{$s$-quark excited states}

At last let us describe excitations in the sector of $s$-quarks. Let us assume that we consider the 1-quark excitation where one quark is taken
from ground state $K=0$ and put to the level for $s$-quark with some total angular momentum $S$. Excited state is $2S+1$ fold
degenerate, we take the mixture \beq |\exc\rangle = \sum_{S_3}\chi_{S_3}|S_3\rangle \eeq where $|S_3\rangle$ are one quark wave functions with
different projections of ${\bm S}$. The calculation of matrix elements gives  now instead of \eq{mat-K}
\beq
\langle S_3|\bm j | S_3'\rangle
=\langle S_3|{\bm S} | S_3'\rangle, \qquad \langle S_3|\frac{\partial \delta M}{\partial\bm\omega}| S_3'\rangle = \zeta_S \langle S_3|{\bm S} |
S_3'\rangle \la{mat-s} \eeq
and matrix elements of $\bm t$ are zero as $s$-quark does not carry isospin. Thus these matrix elements coincide with
\eq{mat-K} for $a_K=0$. The subsequent steps are straightforward. The quantization rule \eq{quant-K} change to
\beq \tilde{\bm T}+\tilde{\bm J}=\hat{\bm S}, \qquad \tilde{Y}=\frac{N_c-3}{3} \la{quant-S}
\eeq
The first rule repeats \eq{quant-K} with evident replacement
$\bm K\to \bm S$. The second rule appears because we substitute the quark with hypercharge $1/3$ (one of $u,d$-quarks in the ground state) by one
$s$-quark (on excited level) with hypercharge $-2/3$. This rule can be also derived directly by calculating coefficient in front of the
Wess-Zumino-Witten term.

Levels of energy for $s$-quark excitations are given by
\[
{\cal E}_S= \frac{c_2(r)-\tilde{T}(\tilde{T}+1)-\frac{3}{4}\tilde{Y}^2}{2I_2}+\frac{1}{2I_1}\left[(1+\zeta_S)\tilde{T}(\tilde{T}+1)-
\right.
\]\beq
\left.
- \zeta_S J(J+1)+\zeta_S(1+\zeta_S){S}({S}+1)\right]
\la{en-S}
\eeq
which is \eq{en-K} with substitution $K\to S$ and $a_K=0$. Available spins are $J=|\tilde{T}-S|\ldots \tilde{T}+S$; collective wave function is analogue of \eq{rotwf-K}:
\[
\Psi_{ S}(R,{\cal S},\chi) = \sqrt{\frac{{\rm dim}(r)(2J+1)}{2S+1}}\times
\]\beq
\times\sum_{\tilde{T},\tilde{T}_3,S_3} C^{SS_3}_{\tilde{T}\tilde{T}_3\;J\tilde{J}_3}
{\cal D}^{(r)}_{\tilde{Y}\tilde{T}\tilde{T}_3;YTT_3}(R^+){\cal D}^{J}_{\tilde{J}_3;J_3}({\cal S}^+)\chi_{S_3}
\la{rotwf-S}
\eeq

To summarize: rotational bands around the given excited intrinsic energy  should be constructed in the following
way. One has to choose $SU(3)$-multiplets which contain states obeying quantization rule for $\tilde{Y}$, read off
the corresponding to this $\tilde Y$ the value of $\tilde{T}$ and use formulae \ur{en-ground}, \ur{en-K}, \ur{en-S} for their rotational energy.

Quark wave functions in the mean field approximation are the product of one-particle wave functions of the filled levels. One has to rotate them according to \eq{rot-of-ferm} and then project to collective wave functions obtained
above (see \eq{rotwf-ground}, \eq{rotwf-K}, \eq{rotwf-S}). ``Projection" means that one has to multiply rotated quark wave function by conjugated
collective wave function and integrate in matrices $R$ and $S$. This will produce quark wave functions of the excited baryons with given quantum numbers.

\section{One quark excitations in the mean field and the quark model}

In the limit of $N_c\to\infty$ the quark model becomes a particular case of the general soliton considered above. Indeed, the mean field
approximation should work for the quark model as well, any inter-quark potential can be considered in the this approximation. As far as we are
discussing only symmetry properties of the quark states, details of the potential are not important.

The real difference between the quark model and the picture  considered above is a symmetry of the mean field. The quark model insists on
the complete spherical symmetry and the only mean field in the quark model is the scalar field $S(\bm x)$. The excitations of the quark
model arise as $SU(6)$-multiplets (to be more precise, multiplets of the $SU(6)\otimes O(3)$, $O(3)$-group standing for orbital angular
momentum). All splittings of such multiplets considered as a small perturbation. This should be confronted to the soliton approach where it is
assumed that the mean field has  symmetry of hedgehog and departure from the $SU(6)$ is not considered to be small, it is
of order of unity even at large $N_c$. Nevertheless we should be able to reproduce multiplets of the quark model taking the spherically
symmetrical mean field.

Due to the Witten quantization rule  the $SU(3)$-multiplets for large $N_c$ becomes large as well. The classification of such multiplets was
developed in \cite{largeNc}.

Every $SU(3)$-multiplet is characterized  by two numbers $p$ and $q$. However, this is inconvenient for our purposes. Let us label multiplets by
i) $\tilde{Y}$ --- what Witten condition is fulfilled by this multiplet, ii) $\tilde{T}$ --- what intrinsic isospin corresponds to this value
and iii) {\em exoticness} $X$ which is defined as a minimal number of quark-antiquark pairs in the wave function of the given baryon. Standard
$p$ and $q$-numbers, dimension of the multiplet and  Casimir operator can be expressed in terms of these numbers as follows: \beq
p=2\tilde{T}-X\geq0, \qquad q=\frac{3}{2}\tilde{Y}+2X-\tilde{T}\geq0 \la{pq-condition} \eeq and
\[
{\rm
dim}\!=\!\frac{2\tilde{T}\!+\!1\!-\!X}{2}\left(\frac{3}{2}\tilde{Y}\!+\!1\!+2\!X\!-\!\tilde{T}\right)\left(\frac{3}{2}\tilde{Y}\!+\!2\!+\!X\!+\!\tilde{T}
\right),\]\beq c_2(r)=\frac{3}{4}\tilde{Y}(\tilde{Y}+2)+\tilde{T}(\tilde{T}+1)+X\left(\frac{3}{2}\tilde{Y}+1-\tilde{T}\right)+X^2 \la{dimcSU3}
\eeq Plugging these expressions into the general formula for the energy \eq{en-K} we arrive at:
\[
{\cal M}_K={\cal M}_0+\Delta{\cal E}  +\frac{1}{2I_1}\left[\tilde{a}_K J(J+1)+(1-\tilde{a}_K)\tilde{T}(\tilde{T}+1)
\right.
\]\beq
\left.
-\tilde{a}_K(1-\tilde{a}_K)K(K+1)
\right]+\frac{(1+X)(2+3\tilde{Y})}{2I_2}
\la{en-SU3}
\eeq
We see that $I_2$ plays the role of the moment of inertia for  exotic states; their
spectrum is equidistant and distances between states are of order  unity (we remind that $I_2\sim N_c$ and $\tilde Y \sim N_c$).
Moment of inertia $I_1$ governs ordinary excitations splitting \cite{largeNc}. We will not consider the {\em rotational} exotics here, exotic states
have some specifics related to the fact that their widths are $\sim O(1)$ \cite{Pra, Diakonov:2008hc}. Anyway, these states are separated from the normal
rotational band by the interval $\sim O(1)$.

Every excited state has a restricted number of {\em non-exotic} states  entering the rotational band with definite $\tilde T$. They are
determined from the condition that $p\geq0$ and $q\geq0$. In particular, for excitations in sector of $s$-quarks at $N_c=3,\tilde{Y}=0$ we get
only one state --- singlet with spins $J=S\pm 1/2$ (where $S$ is the spin of excited states) and other multiplets are exotic. At larger $N_c$
$s$-quark excitations form non-trivial rotational band with energies given by \eq{en-SU3} and $\tilde{a}_K=-\zeta_s$ (see \eq{en-S}).

Excitations in sectors of $u,d$-quarks have richer structure. For non-exotic states ($X=0$) at $N_c=3$ and $\tilde{Y}=1$
it follows from \eq{pq-condition} that we have only two possibilities: $\tilde{T}=\half$ and  $\tilde{T}=\frac{3}{2}$. In other words they can come only as octets or decuplets. At larger $N_c$ other multiplets become non-exotic. At  $K=0$ we obtain the rotational band of $J=\tilde{T}$ with different spins changing in the limits $\frac{1}{2}<J<\frac{N_c}{2}$. Their energies are given by general formula \eq{en-SU3} with $X=0$, $\tilde{a}_K=0$.

At $K=1$ we have 3 possible series of rotational bands $J=\tilde{T}-1,\tilde{T},\tilde{T}+1$ and at $K=2$ there are 5 series with
$J=\tilde{T}-2,\ldots \tilde{T}+2$. Possible spins are determined again from \eq{pq-condition}.

Let us confront this picture to the quark model (see, e.g.,\cite{quark}). The lowest state of the quark model is $56$-plet with orbital moment $L=0$. This multiplet can be generalized to arbitrary $N_c$, its dimension is
\[
{\rm dim}_{'56'}=\frac{1}{120}(N_c+1)(N_c+2)(N_c+3)(N_c+4)(N_c+5),
\]\beq
C_2(SU(6))=\frac{5}{12}N_c(N_c+6)
\la{su6-56}
\eeq
The analogue of '56' in the mean field picture is rotational band around ground state baryons. It is easy to check that
dimension of the full rotational band of ground state baryons:
\[
\sum_{J=\half}^{N_c} (2J+1){\rm dim}(J,X=0,\tilde{Y}=\frac{N_c}{3}) = {\rm dim}_{'56'}
\]
In other words rotational band around ground state completely coincides with '56'-plet. At $N_c=3$  it reduces to the familiar $(\bm{56})=(\bm 8
\frac{\bm 1}{\bm 2})\oplus (\bm 1\bm 0 \frac{\bm 3}{\bm 2})$.

The next $SU(6)$ multiplet is $\bm 7 \bm 0$. Its dimension at arbitrary $N_c$ is equal to:
\[
{\rm dim}_{'70'}=\frac{1}{24}(N_c-1)(N_c+1)(N_c+2)(N_c+3)(N_c+4),
\]\beq
C_2(SU(6))=\frac{1}{12}N_c(5N_c+18)
\la{su6-70}
\eeq
It consists of the following 5 series of $SU(3)$ multiplets:
\begin{enumerate}
\item
3 series of multiplets with $\tilde{Y}=N_c/3$: $\tilde{T}=J-1$  with $J=\frac{3}{2}\ldots\frac{N_c}{2}$, $\tilde{T}=J$ with
$J=\half\ldots\frac{N_c}{2}-1$ and  $\tilde{T}=J+1$ with $J=\half\ldots\frac{N_c}{2}-1$.
\item
2 series of multiplets with $\tilde{Y}=(N_c-3)/3$: $\tilde{T}=J-\half$ with $J=\half\ldots\frac{N_c}{2}-1$ and
$\tilde{T}=J+\half$ with $J=\half\ldots\frac{N_c}{2}-2$
\end{enumerate}
It is easy to check using \eq{dimcSU3} that the total dimension of all 5 series is equal to ${\rm dim}_{'70'}$.

The contents of the $SU(3)$-multiplets entering $'70'$-plet implies that in the mean field picture it consists of one-quark excitation in sector
of $s$-quark with spin $1/2$ and $K=1$ excitation in the sector of $u,d$-quarks. These states become degenerate in the case of spherically
symmetrical mean field. However, there are {\em more} states in the mean field approximation: in this approximation {\em all} states with
$\half\leq \tilde{T}\leq \frac{N_c}{2}$ are present. We will see that absent states are spurious: they are forbidden by the Pauli principle which is
not accounted in the mean field approximation.

'70'-plet is used in the quark model, e.g.,  in order to describe baryons with negative parity \cite{quark}.  It is assumed that these baryons
are described the representation $(\bm 7 \bm 0, \bm 3)$ representation of the $SU(6)\otimes O(3)$: the baryons have
relative orbital angular moment  of two quarks $L=1$. This provides negative parity and makes the total
baryon wave function symmetrical (or antisymmetrical if one accounts for color) under simultaneous exchange of spin, flavor indices ($SU(6)$)
and coordinates. Adding $L=1$ to the mean field multiplets obtained above we see that these baryons should be described by $K=0,1,2$ excitations
in the sector of $u,d$-quarks and $S=\half, \frac{3}{2}$ excitations in the sector of $s$-quarks and their rotational bands. The difference with
quark model is that these 5 sets of states are splitted by $O(1)$ owing to the hedgehog (not spherical) symmetry of the mean field from the very
beginning.

Other multiplets of the quark model can be also analyzed in the same way and one can find their interpretation in
terms of states  appearing in the mean field approximation.

Presented above series of five $SU(3)$ multiplets,  which become at large $N_c$ '70'-plet of $SU(6)$, were observed first in remarkable paper
\cite{Cohen1}. In this paper the approach close in spirit to approach of Manohar et al [27]. was used. Additionally, Authors of
Ref.\cite{Cohen1} claimed a relation between masses of multiplets with different grand spin $K$ which in our notation reads: \beq \Delta{\cal
E}(0\to 0)+3\Delta{\cal E}(0\to 1)+5\Delta{\cal E}(0\to 2)=0 \eeq This relation looks surprisingly since it cannot be fulfilled for an arbitrary
external field. In particular, it is not fulfilled in the exactly solvable model which we consider in Appendix A.

Situation in the sector with  positive parity is completely different. The quark model prediction for excited baryons in this sector consists of
five $SU(6)\otimes O(3)$ multiplets\cite{quark1}: $(\bf {56,0})$ (radial excitation of the ground multiplet), $({\bf 56,2})$, $({\bf 70,0})$,
$({\bf 70,2})$ and  $({\bf 20,1})$. All these multiplets should be nearly degenerate in the quark model (they are degenerate
in the oscillator limit of the model \cite{quark1}). The vast majority of the states in these multiplets is not observed in nature.

Contrary to the situation with negative parity baryons,  some quark model multiplets in the parity-plus
sector have no interpretation as 1-quark excitations in the mean field at $N_c\to\infty$. Instead they correspond to, at least, two-quark
excitations when 2 quarks from the ground level are transferred to the excited levels (possibly with different $K$). It is especially clear for
{({\bf 20,1}) multiplet where $SU(6)$  wave function is completely antisymmetric. This
multiplet cannot fit} the picture where $N_c-1$ quarks are sitting on the same level (and thus have completely symmetric wave function) and only
last quark carries angular momentum (to be more precise, non-zero $K$). The same statement is true for $({\bf 70,0})$ and $({\bf 70,2})$
whose also have mixed symmetry. The 1-quark excited multiple should be symmetric in the $SU(6)$ index of
the excited quark and hence antisymmetric in the first two sitting in the ground state.

One can consider the two-particle excitations in the  mean field at $N_c\to\infty$
as well, keeping in mind, however, that the ${\bf 70}$ and ${\bf20}$ multiplets should be generalized
 to large $N_c$ in different way than it was done for the negative parity baryons. We will not come in
details but we mention that  '20'-plet at arbitrary $N_c$ has dimension $\frac{1}{12}(N_c-2)(N_c-1)(N_c+1)(N_c+2)(N_c+3)$
and corresponds to 10 series of $SU(3)$-multiplets with different
spins ($SU(6)$ Casimir operator is $N_c(5N_c+6)/12$). Dimension of two-quark excited '70'-plet
with positive parity is $\frac{1}{3}(N_c-2)(N_c-1)N_c(N_c+2)(N_c+4)$ and it contains 40 series of $SU(3)$-multiplets  ($SU(6)$
Casimir operator is $3+N_c(5N_c+6)/12$). Of course, most of the states are spurious at $N_c=3$ . One can decompose these series to the state
corresponding to different $K$ and check that they are, indeed, 2-quark excitations (e.g. in the toy model considered in Appendix A).

However, in this paper we would like to insist that even positive parity baryons can be also constructed as one-quark
excitations in the mean field. We believe that baryons with two quarks excited are much heavier and have larger width than one-quark
excitations. In other words, we think that quark model is misleading for the positive parity excited baryons (in particular, $SU(6)$ group is
completely broken) and its obvious success in the sector with negative parity is explained  by the fact that negative parity baryons {\em are}
1-quark excitations. This point of view is supported by the the fact that significant part of the quark model states with positive parity was
never observed. We will adopt the idea that they are described by the same set $K=0,1,2$ levels as for the negative parity
baryons. One can easily construct the wave functions of these states directly and check that there are only one spurious
multiplet among them (for more details, see next Section). Our classification, as was already said, does not correspond to the quark model.

Different situation arises if one puts $N_c=3$ from the very beginning.
Then identification of one-particle and two-particle excitations becomes difficult. Indeed, after separation of the center of
 motion movement, the wave functions are not the product of
one-particle states. Moreover, one can impose  $SU(6)$ limit (central symmetrical scalar mean
field) and then proceed to the oscillator limit of the interaction. Then all constructed $K=0,1,2$ states should fall in one of the five $SU(6)$
multiplets mentioned above (We remind that mean field approximation becomes exact for oscillator model). And they do: one can construct directly
wave functions of these states with  help of the method described at the end of the previous Section and confront them to the $SU(6)$
wave functions built in \cite{quark1} (Appendix B). Then one can see that completely antisymmetric $SU(6)$-multiplet $\bf 20$ remains  2-quark
excitation even at $N_c=3$. The contents of $K=0$ and $K=2$ states is completely exhausted by $(\bf 56,0)$ and $(\bf 56,2)$ multiplets. At last,
all one-quark $K=1$ states are inside the multiplets of mixed symmetry $(\bf 70,0)$ and $(\bf 70,2)$.

To summarize: one-particle excitations in the mean field at large $N_c$ with lowest $K=0,1,2$ is only the small part of quark model  multiplets
for the positive parity. However, as we shall see in the next Section, it is this part which is observed in Nature (at least below 2
GeV).

\section{Comparison with the experimental spectrum}

We shall now look into the experimental spectrum of light baryon resonances up to around
2 GeV, trying to recognize among them the rotational bands about the one-quark excitations
from the ground state to the intrinsic quark levels. We shall treat the quantities
$I_1$,  $\tilde{a}_K$ and $\Delta{\cal E}$ entering  \eq{en-SU3}
as free parameters to be fitted from the known masses, although in principle they are calculable if the
(self-consistent) mean field is known. Still, there are much more resonances than free parameters, therefore
 the rotational bands are severely restricted by \eq{en-SU3} . As we shall see these restrictions are well supported
 by experimental observations, despite that in the real world $N_c$ is only three.

Since at this time we do not take into account the splittings inside $SU(3)$ multiplets as due to the nonzero $m_s$, \eq{en-SU3} should be
compared with the {\it centers} of multiplets. For the octet, the center is defined as ${\cal M}_8=(2m_N+2m_\Xi+3m_\Sigma+m_\Lambda)/8$, and for
the decuplet it is ${\cal M}_{10}=(4m_\Delta+3m_{\Sigma^*}+2m_{\Xi^*}+m_\Omega)/10\approx m_{\Sigma^*}$. We take the phenomenological values for
${\cal M}_8,\,{\cal M}_{10}$ from the paper by Guzey and Polyakov~\cite{GP} who have  analyzed baryon multiplets up to 2
GeV.

\subsection{Spurious states}

Comparing the mean-field predictions (valid at large $N_c$) with the data, it should be kept in mind that certain
rotational states are, in fact, spurious, as they are artifacts of the mean-field approximation where the spatial wave function is a product of
one-particle wave functions. When averaging over the center of mass is taken into account (which is an ${\cal O}(1/N_c)$ effect) the baryon wave
functions depend only on the differences of quark coordinates, which for some states may contradict the Pauli principle. This effect has been
long known both in nuclear physics~\cite{ES} and in the non-relativistic quark model~\cite{quark}. The simplest way to identify spurious states
is to continuously deform the mean field to the non-relativistic oscillator potential where the wave functions are explicit. Again, they can be
written directly projecting rotated mean field quark wave functions with collective wave functions constructed here. If some state is absent in
that limit, it cannot appear from a continuous deformation. An independent way to check for spurious states is to deform the problem at hand to
the exactly solvable (0+1)-dimensional four-fermion interaction model~\cite{Pobylitsa} where the large-$N_c$ approximation is also possible and
reveals extra states (see Appendix A).

Specifically, in the parity-plus sector, the spurious state is $({\bf 10},1/2^+)$ arising from the rotational band about the $(0^+\to 2^+)$
transition. Such state arises also from the $(0^+\to 1^+)$ transition but then it is not spurious.

In the parity-minus sector there are more spurious states: the multiplets
$({\bf 10},5/2^-)$ and $({\bf 10},7/2^-)$ stemming from the $(0^+\to 2^-)$ transition
are spurious, two out of three multiplets $({\bf 10},3/2^-)$ arising from
$(0^+\to 0^-,1^-,2^-)$ transitions are spurious, and one out of two multiplets
$({\bf 10},1/2^-)$ stemming from $(0^+\to 1^-,2^-)$ transitions is spurious, too.
As it was already said, remaining negative parity multiplets exactly coincide with
octets and decuplets from $\bf (70,1)$ multiplet of $SU(6)\otimes O(3)$ of the quark model.

Spurious rotational states should be removed from the consideration.

\subsection{Parity-plus resonances}

The two lowest multiplets $({\bf 8},1/2^+,1152)$ and $({\bf 10},3/2^+,1382)$
(the last number in the parentheses is the center of the multiplet) form the rotational
band about the ground-state filling scheme shown in Fig.~2. Fitting these masses
by \Eq{en-SU3} we find ${\cal M}_0+\frac{3}{4I_2}=1090\,{\rm MeV}$, $1/I_1=153\,{\rm MeV}$.

Apart from the two lowest multiplets, there is another low-lying pair with the same quantum
numbers, $({\bf 8},1/2^+,1608)$ and $({\bf 10},3/2^+,1732)$. Other parity-plus multiplets
are essentially higher. One needs a $0^+\to 0^+$ transition to explain this pair.
Comparing the masses one finds that the second $K^P=0^+$ intrinsic quark level must
be 482 MeV higher than the ground state $0^+$ level, $\Delta{\cal E}(0^+\!\to\!0^+)=482\,{\rm MeV}$.
The moment of inertia appears to be considerably larger than for the ground-state multiplets,
$1/I_1=83\,{\rm MeV}$. Although the difference is an ${\cal O}(1/N_c)$ effect, it may be
enhanced if the radially excited $0^+$ level has a much larger effective radius.

\begin{center}
\begin{table*}[h]
\begin{tabular}{|l|l|c|c|}
\hline
quark levels  & rotational bands& $(I_1)^{-1}$, MeV & $\tilde{a}_K$ \\
\hline
$K^P=0^+$, ground& $({\bf 8},1/2^+,1152)$\quad $({\bf 10},3/2^+,1382)$ & \multirow{2}*{153} &  \\
state &&&\\
\hline
$0^+\to 0^+$\quad 482 MeV & $({\bf 8},1/2^+,1608)$\quad $({\bf 10},3/2^+,1732)$ & 83 &\\
\hline
\multirow{3}*{$0^+\to 2^+$\quad 722 MeV} & $({\bf 8},3/2^+,1865)$\quad $({\bf 8},5/2^+,1873)$&\multirow{3}* {131} &\multirow{3}*{-0.050} \\
&$({\bf 10},3/2^+,2087)$\quad $({\bf 10},5/2^+,2071)$&&\\
& $({\bf 10},7/2^+,2038)$ &&\\
\hline
\multirow{3}*{$0^+\to 1^+\;$ $\sim\!780\,{\rm MeV}$} & $N(1/2^+,1710)$\quad $N(3/2^+,1900)$ &&\\
&  $\Delta(1/2^+,1910)$ \quad $\Delta(3/2^+,\!\sim\!1945)$? && \\
& $\Delta(5/2^+,2000)$ && \\
\hline
\hline
\multirow{2}*{$0^+\to 1^-$\quad 468 MeV} & $({\bf 8},1/2^-,1592)$\quad $({\bf 8},3/2^-,1673)$  &\multirow{2}*{171}&\multirow{2}*{ 0.336}\\
&$({\bf 10},1/2^-,1758)$ \quad $({\bf 10},3/2^-,1850)$&&\\
\hline
$0^+\to 0^-$\quad 563 MeV & $({\bf 8},1/2^-,1716)$&155(fit)& \\
\hline
$0^+\to 2^-$\quad 730  MeV & $({\bf 8},3/2^-,1896)$\quad $({\bf 8},5/2^-,1801)$ & 155(fit) & -0.244 \\
\hline
$0^+\to \frac{1}{2}^-$\quad 254  MeV & $({\bf 1},1/2^-,1405)$&&\\
\hline
$0^+\to \frac{3}{2}^-$\quad 379  MeV & $({\bf 1},1/2^-,1520)$&&\\
\hline
\end{tabular}
\caption{Interpretation of all baryon resonances below 2 GeV, as rotational excitations on top of
intrinsic quark states.}
\end{table*}
\end{center}

There is a group of five multiplets, $({\bf 8},3/2^+,1865)$, $({\bf 8},5/2^+,1873)$, $({\bf 10},3/2^+,2087)$, $({\bf 10},5/2^+,2071)$, $({\bf
10},7/2^+,2038)$ that are good candidates for the rotational band around the $0^+\to 2^+$ transition. Indeed, this is precisely the content of
the rotational band for this transition (the spurious multiplet $({\bf 10},1/2^+)$ excluded), and a fit to the masses according to \eq{en-SU3}
gives a small $\sqrt{\chi^2}=15\,{\rm MeV}$. It should be kept in mind, however, that not all members of all multiplets are well
established~\cite{GP}, and those that are, have an experimental uncertainty in the masses. It means that the `experimental' masses for the
centers of multiplets are known at best to an accuracy of 20-40 MeV. We find from the fit $1/I_1=131\,{\rm MeV}$, $\Delta{\cal E}(0^+\!\to\!
2^+)=722\,{\rm MeV}$. Therefore, the intrinsic $2^+$ level must be higher than the $0^+$ one.

The only relatively well established multiplet that is left in the range below 2 GeV,
is $({\bf 8},1/2^+,1846)$. It prompts that it can arise from the rotational band about
the $0^+\to 1^+$ transition, however, other parts of the band are poorly known.
If one looks into non-strange baryons that are left, one finds $N(1/2^+,1710^{***})$,
$N(1/2^+,1900^{**})$, $\Delta(1/2^+,1910^{***})$ and $\Delta(5/2^+,2000^{**})$, with $\Delta(3/2^+)$
missing. The quantum numbers and the masses of these supposed resonances fit rather well
the hypothesis that they arise as a rotational band about the $0^+\to 1^+$ transition, however,
their low status prevents a definite conclusion. The intrinsic $1^+$ level must be approximately
60 MeV higher than the $2^+$ quark level.

\subsection{Parity-minus resonances}

The situation here is similar to the parity-plus sector: one needs intrinsic quark levels
with $K^P=0^-,1^-,2^-$ to explain the resonances as belonging to rotational bands about
these transitions. Given that several rotational states in the parity-minus sector are
spurious, one expects to find the following multiplets stemming from these transitions:
$({\bf 8},1/2^-)\times 2$, $({\bf 8},3/2^-)\times 2$, $({\bf 8},5/2^-)$,
$({\bf 10},1/2^-)$, $({\bf 10},3/2^-)$: these are precisely the observed multiplets.

We know that all remaining multiplet are spurious but we do not know the way to assign specific
$K$ to the observed one. We attribute them according to \eq{en-SU3} requiring that no mixing
can happen ($\zeta_K=0$). There is only one way to do this.

We assign 4 lowest multiplets $({\bf 8},1/2^-,1592)$, $({\bf 8},3/2^-,1673)$, $({\bf 10},1/2^-,1758)$ and $({\bf 10},3/2^-,1850)$ to the
rotational band of $K=1^-$ level. The fit tells that corresponding moment of inertia is $1/I_1=171\,{\rm MeV}$ and energy of the level
$\Delta{\cal E}(0^+\!\to\!1^-)=468\,{\rm MeV}$ is close  to $\Delta{\cal
E}(0^+\!\to\!0^+)$. This does not look impossible.

The multiplet $({\bf 8},1/2^-,1716)$ should be ascribed as $0^+\to 0^-$ transition and two remaining multiplets
$({\bf 8},3/2^-,1896)$ and $({\bf 8},5/2^-,1801)$ to $0^+\to 2^-$ transition.

These assignments produce reasonable values of mixing coefficients $\tilde{a}_K$ which can be explained without mixing of
rotations and other degrees of freedom in the effective meson Lagrangian. Probably, some other information(mass splittings or
resonance widths) should be used to fix finally the attribution of multiplets to the rotational bands. If the final scheme
would be different from assumed here, it will witness the large role of other than pion mesons in formation of negative
parity baryons.

To summarize, all parity-plus and parity-minus baryons around 2 GeV and below can be accommodated
by the scheme, assuming they all arise as rotational excitations about the $0^+\to 0^+,1^+,2^+$
and $0^+\to 0^-,1^-,2^-$ transitions, see Table~1. There are no unexplained resonances left,
but there appears an extra state $\Delta(3/2^+,\!\sim\!\!1945)$ stemming from the $0^+\to 1^+$
transition, which is so far unobserved, so this state is a prediction.

\subsection{$s$ quarks}

As emphasized in Section 3, $s$ quarks are in a completely different external field than $u,d$
quarks, even in the chiral limit. Only the confining forces which we model by a linear rising
scalar field are the same for all quarks. The two excited levels for $s$ quarks are shown in
Fig.~3: they are needed to explain the singlet $\Lambda(1/2^-,1405)$ and $\Lambda(3/2^-,1520)$
resonances. The corresponding values of $\Delta \cal{E}$ presented in the Table 1.
 No more singlet $\Lambda$'s are known below 2 GeV, therefore there should be no
intrinsic $s$-quark levels either with positive or negative parity in this range.

Following the standard logic of the quark model we describe in this paper all baryon resonances as excitations of {\em valence quarks} (see
Fig.\ref{fig:2}). This is not necessary, however. New resonances can appear due to transitions from the levels which belong to the Dirac
continuum. The main configuration for such baryons will consist of 5 quarks (3 valence quarks plus quark-antiquark pair) but this does not mean
that they should be exotic. Just opposite, most of them would be ordinary octets and decuplets.

\begin{figure}[h]
\begin{minipage}[]{0.42\textwidth}
\hspace{4cm}
\includegraphics[width=\textwidth]{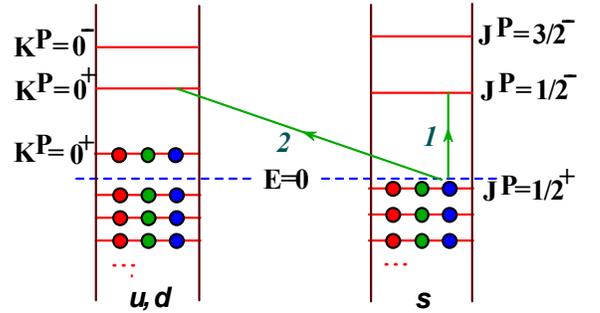}
\end{minipage}
\caption{(Color online) Possible transitions of the $s$ quark from the highest filled $s$ level to excited $s$
levels ({\it 1}), and to excited $u,d$ levels ({\it 2}).}
\label{fig:4}
\end{figure}

For example, the intriguing question is where is the highest {\em filled} level of $s$ quarks? Presumably, it must be a level with quantum
numbers $J^P=\half^+$ as possessing maximal symmetry. There can be one-quark excitations from that level both to the $s$-quark excited levels
$\frac{1}{2}^-$ and $\frac{3}{2}^-$, and to the $u,d$ excited levels $0^+,0^-,...$, see Fig.\ref{fig:4}. Transitions of the first type generate
a rotational band consisting of $({\bf 8},1/2^-)\times 2$, $({\bf 8},1/2^-)$, $({\bf 10},1/2^-)$, $({\bf 10},3/2^-)\times 2$ and $({\bf
10},5/2^-)$. Transitions of the second type called, in the terminology of nuclear physics, Gamov-Teller transitions, generate the exotic {\em
antidecuplet} $({\overline{\bf 10}},1/2^+)$, etc.~\cite{Diakonov:2009kw} . It turns out that it is difficult, if not impossible, to move the
highest filled level of $s$ quarks $\half^+$, which must satisfy \Eq{s_P=J-} in a `realistic' mean field, more than $\sim 700\,{\rm MeV}$ below
the first excited level $\half^-$. Therefore, the parity-minus resonances generated by the transition {\it 1} in Fig.~4 must reveal themselves
in the spectrum below 2 GeV. We note that such resonances will have a substantial 5-quark component $u(d)u(d)u(d)s\bar s$ since they require an
$s$ quark to be pulled out of the filled level and put onto an excited level. Probably, the real-world resonances are certain mixtures of these
excitations with the $u,d$ excitations described in the previous section. This is a welcome feature as, for example, the well-known resonance
$N(1/2^-,1535)$ has a surprisingly large coupling to the $\eta$ meson (see also \cite{Cohen1,Zou}). The `Gamov-Teller' transition 
{\it 2} gives a natural explanation of the exotic $\Theta^+$ resonance~\cite{DPP-97} exactly at the position where it has been claimed by a number of
experiments~\cite{Diakonov:2009kw} .

\section{Mass splittings}
The non-zero mass of the strange quark $m_s$ breaks down $SU(3)$ flavor group and splits $SU(3)$ multiplets. Let us calculate these
splittings. Inserting quark mass $m$ (a matrix in flavor) into the quark determinant \eq{SeffRot} and expanding up to the first order in $m$ we
obtain
\[
\delta_m S=-i\sum_c{\rm Sp}_{\text{occupied}}\biggl\{R^+mR\gamma^0\times
\]\beq
\times \left. \frac{1}{i\frac{\partial}{\partial t}-{\cal H}(M+\delta M)-\tOmega t_a-\tomega_i j_i}\right\}
\eeq
Mass of strange quark has both
singlet and octet part $m=m_0{\bm 1}+m_8\lambda_8$ and $m_8=m_s/\sqrt{3}$ and splittings are determined only by octet part $m_8$.

We want to calculate mass splitting in the zeroth and  the first order in angular and flavor frequencies $\tomega$ and $\tOmega$. For ground state baryons this calculation was carried out in many papers (see, e.g. \cite{Rev}).  Result reads
\[
\delta_m S = \frac{m_s}{\sqrt{3}}\int\! dt\;\biggl[\sum_a\sum_{\text{occup}}{\cal D}^{(8)}_{8a}(R) \langle n|\lambda_a\gamma^0|n\rangle+
\]\beq
\left. +2K_1 \sum_{i=1}^3 {\cal D}^{(8)}_{8i}(R)(\tOmega_i-\tomega_i)+ 2K_2 \sum_{a=4}^7 {\cal D}^{(8)}_{8a}(R)\tOmega_a \right] \la{ms-act}
\eeq
Here the first term is of the zeroth order in frequencies, the second and the third terms represent the first order
corrections. $K_1$ and $K_2$ are some constants analogous to the moments of inertia \eq{Inglis-inertia} (again $n$ runs over occupied
levels and $m$ runs over free levels in the mean field):
\beq K_{ab}\!=\!N_c\!\sum_{n,m}\! \frac{\langle n|\gamma^0 t_a|m\rangle
\langle m|\gamma^0 t_b|n\rangle\!+\!\langle n|\gamma^0 t_b|m\rangle \langle m|\gamma^0t_a|n\rangle}{\varepsilon_m-\varepsilon_n}
\la{K-def}
 \eeq
(if there is a mixing of rotations with $\delta M$ this expression should be modified). Tensor $K_{ab}$ has a structure analogous to the structure of the
moment of inertia
\beq K_{ab}= \left\{
\begin{array}{ll}
K_1\delta_{ab} &{a,b=1\ldots 3} \\
K_2\delta_{ab} &{a,b=4\ldots 7} \\
0, & a=b=8\\
\end{array}
\right.
\eeq

Expression \ur{ms-act} is valid for rotational bands above the ground state and one-particle excitations. First term for ground state baryons is non-zero only at $a=8$, it can be expressed through the experimentally known quantity --- so-called $\Sigma$-term:
\[
\sum_c\sum_{\text{occup}}\langle n|\lambda_8\gamma^0|n\rangle = \frac{1}{3}\frac{m_s}{m_u+m_d}\Sigma,
\]\beq
\Sigma =(m_u+m_d) \frac{\partial{\cal M}}{\partial(m_u+m_d)} \eeq
It was used here that all valence levels are
located in the sector of $u,d$-quarks. We will imply below only this case. Indeed, we have seen that at $N_c=3$ one-particle excitations in the
sector of $s$-quarks (from the ground state) are singlets, so there is no mass splitting  present for this type of excitations.

In the sector of $u,d$-quarks there is also another possibility $a=1,2,3$ for excited level:
\beq
\langle \exc|\frac{\lambda_i}{2}\gamma^0|\exc\rangle =d_K\sum_{K_3,K_3'}
\chi^+_{K'_3}\chi_{K_3}\langle K'_3| K_i | K_3\rangle ,
\eeq
where $d_K$ is some constant which is determined by wave function
\beq
d_K=\pm \int\! dr r^2 \left[\frac{g^2(r)-f^2(r)}{2K}-\frac{h^2(r)-j^2(r)}{2(K+1)}\right]
\eeq
where plus sign stands for the states with parity $(-1)^K$ and minus for states with parity $(-1)^{K+1}$. Calculation of $d_K$
is analogous to the calculation of coefficient $c_K$  (see above).

First order terms in frequencies in \eq{ms-act} can be simplified as well. We  substitute frequencies by operators
$\tilde T_a$ according to \eq{forms}. Using relation:
\[
T_8= \sum_{a=1}^8 {\cal D}^{(8)}_{8i}(R)\tilde{T}_a
\]
one can express the last term in \eq{ms-act} in terms of the sum with $a=1,2,3$ and hypercharge $Y=2T_8/\sqrt{3}$.
Proceeding to the Hamiltonian
\[
{\cal H}_m= \alpha {\cal D}^{(8)}_{88}(R) + \beta Y + \sqrt{3}\gamma \sum_{i=1}^3{\cal D}^{(8)}_{8i}(R)\tilde{T}_i+
\]\beq
+\sqrt{3}\delta\sum_{i=1}^3 {\cal D}^{(8)}_{8i}(R)\hat{K}_i
\la{ms-ham}
\eeq
where
\[
\alpha=-\frac{2}{3}\frac{m_s}{m_u+m_d}\Sigma+m_s\frac{K_2}{I_2}, \quad
\beta=-m_s\frac{K_2}{I_2},
\]\beq
\gamma=\frac{2m_s}{3}\left(\frac{K_1}{I_1}-\frac{K_2}{I_2}\right), \quad
\delta=\frac{2m_s}{3}\left(d_K-\frac{K_1}{I_1}\tilde{a}_K\right)
\eeq

We see that mass splittings are determined by four possible structures. Only the last term is novel, other three are known for ground state baryons. Moreover, constants $\alpha,\beta,\gamma$ up to corrections of order $1/N_c$ are the same for all levels. As to $\delta$ it is determined by the properties of the excited level and is individual for given level. Nevertheless, $\delta$ is the same for all rotational band of the given level
Note also that $\alpha\sim O(N_c)$ while $\beta,\gamma,\delta\sim O(1)$.

Mass splittings are determined by the average of the Hamiltonian \eq{ms-ham} in collective wave functions \eq{rotwf-ground} and \eq{rotwf-K}. Resulting expressions, of
course, respect Gell-Mann-Okubo formula. We parameterize the masses of particles in the octet as:
\[
{\cal M}_{N}=M_8-\frac{7}{4}\mu^{(8)}_1-\mu^{(8)}_2, \quad  {\cal M}_{\Lambda}=M_8-\mu^{(8)}_1,
\]
\beq
{\cal M}_{\Sigma}=M_8+\mu^{(8)}_1 \quad {\cal M}_{\Xi}
=M_8+\frac{3}{4}\mu^{(8)}_1+\mu^{(8)}_2
\eeq
and masses of decuplet particles as
\[
{\cal M}_{\Delta}=M_{10}-\mu^{(10)}, \quad  {\cal M}_{\Sigma}=M_{10}, \qquad {\cal M}_{\Xi}=M_{10}+\mu^{(10)}
\]\beq
{\cal M}_{\Omega}
=M_{10}+2\mu^{(10)}
\eeq
This parametrization obeys Gell-Mann-Okubo formula automatically. In the Table 2 we give the values of $\mu$ in terms of $\alpha,\beta,\gamma, \delta$ for different
values of $K$ and the spin of the multiplet $J$.
\begin{center}
\begin{table*}[ht]
\begin{tabular}{|c|c|c|c|c|c|}
\hline
K & Rep & J & $\mu_1^{(8)}$ & $\mu^{(8)}_2$ &$\mu^{(10)}$ \\
\hline
\multirow{2}*{0}&8&$\half$ & $-\frac{\alpha}{10}-\frac{3\gamma}{20}$&$-\frac{\alpha}{8}-\beta+\frac{5\gamma}{16}$ & \\
\cline{2-6}
& 10&$\frac{3}{2}$& && $-\frac{\alpha}{8}-\beta+\frac{5\gamma}{16}$\\
\hline
\multirow{5}*{1}&\multirow{2}*{8}&$\frac{1}{2}$&$-\frac{\alpha}{10}-\frac{11\gamma}{20}-\frac{3\delta}{5}$&
$-\frac{\alpha}{8}-\beta+\frac{55\gamma}{48}-\frac{5\delta}{4}$&\\
\cline{3-6}
&&$\frac{3}{2}$&$-\frac{\alpha}{10}+\frac{\gamma}{20}+\frac{3\delta}{10}$&
$-\frac{\alpha}{8}-\beta-\frac{5\gamma}{48}-\frac{5\delta}{8}$ &  \\
\cline{2-6}
&\multirow{3}*{10}&$\frac{1}{2}$
&&&$-\frac{\alpha}{8}-\beta+\frac{35\gamma}{48}+\frac{5\delta}{8}$ \\
\cline{3-6}
&&$\frac{3}{2}$&&&
$-\frac{\alpha}{8}-\beta+\frac{23\gamma}{48}+\frac{\delta}{4}$ \\
\cline{3-6}
&&$\frac{5}{2}$&&&
$-\frac{\alpha}{8}-\beta+\frac{\gamma}{16}-\frac{3\delta}{8}$ \\
\hline
\multirow{6}*{2}&\multirow{2}*{8}&$\frac{3}{2}$&
$-\frac{\alpha}{10}-\frac{3\gamma}{4}-\frac{9\delta}{10}$&
$-\frac{\alpha}{8}-\beta+\frac{25\gamma}{16}+\frac{15\delta}{8}$&\\
\cline{3-6}
&&$\frac{5}{2}$&$-\frac{\alpha}{10}+\frac{\gamma}{4}+\frac{3\delta}{5}$&
$-\frac{\alpha}{8}-\beta-\frac{25\gamma}{48}-\frac{5\delta}{4}$ &  \\
\cline{2-6}
&\multirow{4}*{10}&$\frac{1}{2}$
&&&$-\frac{\alpha}{8}-\beta+\frac{17\gamma}{16}+\frac{9\delta}{8}$ \\
\cline{3-6}
&&$\frac{3}{2}$&&&
$-\frac{\alpha}{8}-\beta+\frac{13\gamma}{16}+\frac{3\delta}{4}$ \\
\cline{3-6}
&&$\frac{5}{2}$&&&
$-\frac{\alpha}{8}-\beta+\frac{19\gamma}{48}+\frac{\delta}{8}$ \\
\cline{3-6}
&&$\frac{7}{2}$&&&
$-\frac{\alpha}{8}-\beta-\frac{3\gamma}{16}-\frac{3\delta}{4}$ \\
\hline
\end{tabular}
\caption{Mass splittings for octet and decuplet particles for different $K$.}
\end{table*}
\end{center}
Expressions for mass splittings give rise to the number of relations between masses of the particles entering the same rotational band. These relations are similar to well-known Guadagnini relation which is valid for the ground state octet and decuplet (see, e.g., \cite{DPP-97}). Let us itemize these relations
for $K=2$ rotational band of the baryons with positive parity, for octets and decuplets:
\[
5\mu^{(8)}_2\left(\frac{3}{2}\right)+9\mu^{(10)}\left(\frac{5}{2}\right)=14\mu^{(10)}\left(\frac{3}{2}\right),
\]\beq
5\mu^{(8)}_2\left(\frac{5}{2}\right)+11\mu^{(10)}\left(\frac{3}{2}\right)=16\mu^{(10)}\left(\frac{5}{2}\right),
\eeq
and for decuplets only
\[
5\mu^{(10)}\left(\frac{7}{2}\right)+7\mu^{(10)}\left(\frac{3}{2}\right)=12\mu^{(10)}\left(\frac{5}{2}\right),
\]\beq
3\mu^{(10)}\left(\frac{5}{2}\right)+5\mu^{(10)}\left(\frac{1}{2}\right)=8\mu^{(10)}\left(\frac{3}{2}\right)
\eeq
(we put in parenthesis the spin of the particles). All these relations work with accuracy better than 10\% and some even with accuracy 1-2\%.

For $K=1$ (negative parity) we get two relations:
\[
7\mu^{(10)}\left(\frac{1}{2}\right)+3\mu^{(8)}_2\left(\frac{3}{2}\right)=10\mu^{(10)}\left(\frac{3}{2}\right),
\]\beq
5\mu^{(10)}\left(\frac{3}{2}\right)+3\mu^{(8)}_2\left(\frac{1}{2}\right)=8\mu^{(10)}\left(\frac{1}{2}\right) \eeq While the first is fulfilled
with accuracy 2\%, the second is fulfilled only at 10\% level.

Last relation for $K=0$ (which is precisely Guadagnini's one but for excited baryons) reads: $\mu^{(10)}\left(\frac{3}{2}\right)=\mu^{(8)}_2
\left(\frac{1}{2}\right)$. This relation which is working rather good for ground state octet and decuplet is broken surprisingly strong
for $K=0^+$ excited state.

The situation changes in the strict limit $N_c\to\infty$, in the approach advocated in \cite{DJM1}. According to the last
approach  one should consider Clebsch-Gordan coefficients in the same limit $N_c\to\infty$ . The required isoscalar factors are collected in Appendix \ref{Cle-Gor}, so calculations are straightforward.

The recalculated results demonstrate different $N_c$ counting.  It appears that mass splittings are not $O(m_sN_c)$ but only
$O(m_s)$. Both constants $\alpha$ and $\beta$ enter the leading term, while $\gamma$ and $\delta$ appear in corrections $O(m_s/N_c)$. Probably,
this picture is more satisfactory from the general point of view. Let us note that it coincides with $N_c$ counting developed in
\cite{DJM1,DJM2} and all derived there mass relations are also automatically fulfilled.

Gell-Mann Okubo relations appear to be still valid. This is not trivial, especially for ``decuplets", where not one but two final states at arbitrary $N_c$ are available (so it is possible to talk about $F$- and $D$-scheme for ``decuplets"). However, at large $N_c$ Gell-Mann-Okubo relations are restored, up to the order $O(1/N_c)$ inclusive (they are not exact in $N_c$!). To save space we will not fill up the complete table of masses analogous to the table at $N_c=3$. Instead we write down only mass relations which are independent on the concrete model (some part of them were already known). For $K=0$
\beq
\mu^{(10)}\left(\frac{3}{2}\right)=\mu^{(8)}_2\left(\frac{1}{2}\right)-\frac{1}{4}\mu^{(8)}_1\left(\frac{1}{2}\right)
\eeq
which substitutes Guadagnini's relation  derived at $N_c=3$ (see above). We see that accuracy of this relation is less than of original one. It is not surprising, as the continuation of the Clebsch-Gordan coefficient introduces a new source of inaccuracy. At $K=1$ there are following relations:
\[
12\mu_2^{(8)}\left(\frac{3}{2}\right)-3\mu_1^{(8)}\left(\frac{3}{2}\right)+14\mu^{(10)}\left(\frac{3}{2}\right)
=26\mu^{(10)}\left(\frac{1}{2}\right),
\]
\beq
12\mu_2^{(8)}\left(\frac{1}{2}\right)-3\mu_1^{(8)}\left(\frac{1}{2}\right)+20\mu^{(10)}\left(\frac{3}{2}\right)
=32\mu^{(10)}\left(\frac{1}{2}\right)
\eeq
And at last for $K=2$
\[
20\mu_2^{(8)}\left(\frac{5}{2}\right)-5\mu_1^{(8)}\left(\frac{5}{2}\right)+44\mu^{(10)}\left(\frac{3}{2}\right)
=64\mu^{(10)}\left(\frac{5}{2}\right)
\]\beq
20\mu_2^{(8)}\left(\frac{3}{2}\right)-5\mu_1^{(8)}\left(\frac{3}{2}\right)+34\mu^{(10)}\left(\frac{3}{2}\right)
=54\mu^{(10)}\left(\frac{5}{2}\right)
\eeq
(relation for decuplets is the same as at $N_c=3$). These relations are valid in the linear order in $m_s$ and
up to the order $O(1/N_c)$ inclusive. In general, they are obeyed worse than original ones at $N_c=3$.

\section{Conclusions}

If the number of colors $N_c$ is treated as a free algebraic parameter, baryon resonances are classified in a simple way. At large $N_c$ all
baryon resonances are basically determined by the intrinsic quark spectrum which takes certain limiting shape at $N_c\to\infty$. This spectrum
is the same for light baryons ($q\ldots qq$ with $N_c$ light quarks $q$) and for heavy baryons ( $q\ldots qQ$ with $N_c\!-\!1$ light quarks and
one heavy quark $Q$), since the difference is a $1/N_c$ effect \cite{Jenkins:1996de}.

One can excite quark levels in various ways called either one-particle or particle-hole excitations; in both
cases the excitation energy is ${\cal O}(1)$. On top of each one-quark or quark-antiquark excitation there
is generically a band of $SU(3)$ multiplets of baryon resonances, that are rotational states of a baryon as
a whole. Therefore, the splitting between multiplets is ${\cal O}(1/N_c)$. The rotational band is terminated
when the rotational energy reaches ${\cal O}(1)$.

In reality $N_c$ is only 3, and the above idealistic hierarchy of scales is somewhat blurred. Nevertheless, an inspection of the spectrum of
baryon resonances reveals certain hierarchy schematically summarized as follows:
\begin{itemize}
\item Baryon mass: ${\cal O}(N_c)$, numerically 1200 MeV, the average mass of the ground-state octet
\item One-quark and particle-hole excitations in the intrinsic spectrum: ${\cal O}(1)$, typically 400 MeV,
for example the excitation of the Roper resonance
\item Splitting between the centers of $SU(3)$ multiplets arising as rotational excitations of a given intrinsic state:
${\cal O}(1/N_c)$, typically 133 MeV
\item Splitting between the centers of rotational multiplets differing by spin, that are degenerate in the leading order:
${\cal O}(1/N_c^2)$, typically 44 MeV
\item Splitting inside a given multiplet owing to the nonzero strange quark mass: ${\cal O}(m_sN_c)$, typically 140 MeV.
\end{itemize}

In practical terms, we have shown that all baryon resonances up to 2 GeV made of light quarks can be
understood as rotational excitations about certain transitions between intrinsic quark levels. The quantum numbers
of the resonances and the splittings between multiplets belonging to the same rotational band are dictated
by the quantum numbers of the intrinsic quark levels, and appear to be in good accordance with the data.
The content and the splitting of the lowest charmed (and bottom) baryon multiplets are also in accordance
with their interpretation as a rotational band about the ground-state filling scheme.

In this paper, we have concentrated on the algebraic aspect of the problem leaving aside the dynamical aspects. Dynamical models should answer
the question why the intrinsic quark levels for $u,d$ quarks with $K^P=0^\pm,1^\pm,2^\pm$ and the $s$ quark levels with
$J^P=\frac{1}{2}^\pm,\frac{3}{2}^\pm$, {\it etc.}, have the particular energies summarized in Table 1. However, we feel that it is anyway a step
forward: Instead of explaining two hundreds resonances one needs now to explain the positions of only a few intrinsic quark levels. Fig.~1
illustrates that approximately the needed intrinsic spectrum can be achieved from a reasonable set of mean fields \cite{DPV2012}.

The proposed scheme for understanding baryon resonances has numerous phenomenological consequences
that can be investigated even before real dynamics is considered. Namely, the fact that certain groups
of $SU(3)$ multiplets belong to the same rotational band related to one and the same one-quark transition
implies relations between their couplings, form factors, splittings inside multiplets owing to the nonzero
$m_s$, and so on. \\

\begin{acknowledgements}
The work of D.D. and V.P. is supported partly by Deutsche Forschungsgemeinschaft (DFG)
and partly by Russian Government grants RSGSS-4801.2012.2. A.V. is supported in part by the European Community-Research
Infrastructure Integrating Activity Study of Strongly Interacting Matter" (HadronPhysics3, Grant Agreement No. 28
3286) and the Swedish Research Council grants 621-2011-5080 and 621-2010-3326.

We would like to thank K.Goeke, M.Polyakov, M.Praszalowich, W.Plessas and especially P.V.Pobylitsa for very useful discussions.
\end{acknowledgements}

\appendix%
\section{Toy model}
Our approach can be illustrated by the simple but still instructive model suggested in \cite{Pobylitsa}. The model
is one of the class considered first in context of the nuclear physics in \cite{Lipkin} and is exactly solvable.
It describes 0-dimensional quarks with spin, flavor and color which interact by means of 4-fermion color
blind potential. We consider the specific case of potential and write the Lagrangian of the model in already
bosonized form:
\beq
{\cal L}= \frac{\gamma}{2N_c}\left(\rho^a_i\right) + \psi^+(i\partial_t -\gamma\rho^a_i \lambda^a\sigma_i )\psi
\la{L-toy}
\eeq
Here $\sigma_i$ are Pauli matrices acting on the spin indices of quarks,$\lambda^a$ --- Gell-Mann matrices from
$SU(3)_{flavor}$ group; the sum in color indices is implied. This model is of the type
considered in the main text, in the limit $N_c\to\infty$ it can be considered in the mean field approximation. The symmetry
of $\rho^a_i$ is analogous to the one of the octet of vector mesons.

Integrating over $\rho^a_i$ we arrive at the Lagrangian with four-fermion interaction:
\[
{\cal L}_f= \psi^+i\partial_t\psi+
\frac{\gamma}{2N_c}\left(\psi^+\lambda^a\sigma_i \psi\right)\left(\psi^+\lambda^a\sigma_i\psi\right), \qquad
\]\beq
{\cal H}=-
\frac{\gamma}{2N_c}\left(\psi^+\lambda^a\sigma_i \psi\right)\left(\psi^+\lambda^a\sigma_i\psi\right)
\la{fermion-toy}
\eeq
which has $SU(3)_{flavor}\otimes SU(2)_{spin}$ symmetry.

It is convenient to unite spin and flavor indices in the one $SU(6)$ index and classify states of the model as $SU(6)$-multiplets. However,
these multiplets are split by the potential which is  not  $SU(6)$ symmetric. Let us introduce generators ${\cal T}_a$ of $SU(6)$ group and
generators $T_a$ ($S$) for $SU(3)_{flavor}$ ($SU(2)_{spin}$) group \beq {\cal T}_a =\half \psi^+\Lambda_a\psi, \qquad  T_a =\half
\psi^+\lambda_a\psi, \qquad S_i = \half\psi^+\sigma_i\psi \eeq where $\Lambda_a$ ($a=1\ldots 35$) are Gell-Mann matrices for $SU(6)$. Using the
Fierz identities for $SU(6), SU(3), SU(2)$  group the Hamitonian \eq{fermion-toy} can be identically rewritten as: \beq {\cal H}_f=
\frac{\gamma}{N_c}\left[4 ({\cal T}_a)^2-2(T_a)^2-\frac{4}{3}(S_i)^2\right] \la{Ham-toy} \eeq The first term contains the Casimir operator for
$SU(6)$ group, second for $SU(3)_{flavor}$ and third for $SU(2)_{spin}$ one.

According to Pauli principle the allowed colorless states for $N_c$ quarks should be completely symmetric under exchange of $SU(6)$ indices. There is only one $SU(6)$-multiplet which obeys this condition --- symmetric spinor of $N_c$-th rank. Its dimension is given by \eq{su6-56} of the main text (56-plet at $N_c=3$). The $SU(3)_{flavor}\otimes SU(2)_{spin}$ contents of '56'-plet is discussed in the text and  consists of the series of multiplets with $S=1/2\ldots N_c/2$. Their energies are given by \eq{Ham-toy}
\beq
{\cal E}_{`56`}=\gamma\left( \frac{3N_c}{2}+9-\frac{10}{3N_c}S(S+1) \right)
\la{en56-toy}
\eeq
(we used here expression \eq{su6-56} for $C_2$ and \eq{dimcSU3} for $c_2$ with $\tilde{Y}=N_c/3, X=0,\tilde{T}=J$). The first term here is the classical energy, the second is  the quantum correction and the third is the rotational energy. These formulae coincide with ones obtained in \cite{Pobylitsa} for more general case.

The states \eq{en56-toy} exhaust the spectrum of the model \cite{Pobylitsa} only because this model is too poor. One can easily  generalize the model adding to quarks some internal parameters (indices) and assuming that potential remains the same and does not depend on this "hidden" parameters. In this case Pauli principle does not dictate unique symmetry wave function. In generalized model flavor-spin wave function can have any symmetry in such a way that its product with wave function of "hidden" parameters is totally symmetric as required.

The first excited state of the model corresponds to the $SU(6)$-multiplet with all $SU(6)$-indices completely symmetric except one pair which is
antisymmetric. Its dimension is determined by \eq{su6-70} of the text; at $N_c=3$ it corresponds to 70-plet. Using once more expressions for
$SU(6)$ and $SU(3)$ Casimir operators we obtain from \eq{Ham-toy} \beq {\cal E}^{(1-3)}_{`70`}=\gamma\left[
\frac{3N_c}{2}+5-\frac{4}{3N_c}S(S+1)-\frac{2}{N_c}\tilde{T}(\tilde{T}+1)\right] \la{70ud-toy} \eeq where for 3 different series of rotational
excitations $\tilde T=S-1,S,S+1$. These 3 series correspond, as we shall see, excitations in the sector of $u,d$-quarks. There are also two
series with energies:
 \beq {\cal E}^{(4-5)}_{`70`}\!=\!\gamma\left[
\frac{3N_c}{2}+6-\frac{4}{3N_c}S(S+1)-\frac{2}{N_c}\tilde{T}(\tilde{T}+1)+\frac{3}{2N_c}\right] \la{70s-toy} \eeq
where $\tilde{T}=S\pm \half$.
We recognize exactly those 5 series of $SU(3)\otimes SU(2)$ which were described in the text. Other excited states of the model can be also
constructed.

Now we are going to reproduce eqs.\urs{70ud-toy}{70s-toy} in the mean field approximation. We solve Dirac equation
and the consistency equation following from the Lagrangian \eq{L-toy}
\beq
\rho^a_i\sigma_i\lambda^a\phi=\varepsilon\phi, \qquad \rho^a_i=\phi^+_{\text{ground}}\sigma_i\lambda^a\phi_{\text{ground}}
\eeq
We are looking for the mean field  as $SU(2)$ ''hedgehog'':  $\rho^a_i=\bar\rho\delta^a_i$ for $a=1\ldots 3$ and zero otherwise. There are 3 solutions of the Dirac equation:
\[
\phi^{\alpha i}_0=\left(
\begin{array}{c}
\varepsilon^{\alpha i}\\
0
\end{array}
\right),
\qquad
\phi^{\alpha i}_{1(a)}=\frac{1}{\sqrt{2}}\left(
\begin{array}{c}
\varepsilon^{\alpha j}(\sigma_a)^i_j\\
0
\end{array}
\right),
\]\beq
\qquad
\phi^{\alpha i}_{s(a)}=\frac{1}{\sqrt{2}}\left(
\begin{array}{c}
0\\
\delta^{ia}
\end{array}
\right),
\eeq
Here  $\phi^{\alpha i}_0$ ($\alpha$ -isospinor, $i$ --- spinor index) is the wave function with $K=0$ and energy
$\varepsilon_0=-3\gamma\bar{\rho}$, $\phi^{\alpha i}_{1(a)}$ are (three degenerate)wave function of the  states with $K=1$ and energy $\varepsilon_1=\gamma\bar{\rho}$ in sector of $u,d$ quarks. At last $\phi^{\alpha i}_{s(a)}$  is wave function of (doubly degenerate) level in the sector of $s$-quarks with energy $\varepsilon_s=0$.

We obtain the ground state filling level $\varepsilon_0$ by $N_c$ quarks, then consistency equation gives $\bar{\rho}=-1$. The classical part of the mass (proportional to $N_c$) is obtained  substituting the
ground state wave function to the Hamiltonian \eq{fermion-toy}. The $O(1)$ part of energy implies the calculation
of quantum corrections (it comes from one-loop diagram in the quantum field $\rho^a_i$) and is not interesting for
us (see \cite{Pobylitsa}). However the difference between energy of the ground state and excited ones is calculable and indeed is determined by the difference of 1-quark levels:
\bea
{\cal E}^{(1-3)}_{`70`}-{\cal E}_{`56`}&=&\varepsilon_1-\varepsilon_0+O(\frac{1}{N_c})=-4\gamma
\nonumber\\
{\cal E}^{(4-5)}_{`70`}-{\cal E}_{`56`}&=&\varepsilon_s-\varepsilon_0+O(\frac{1}{N_c})=-3\gamma
\eea
(we have to choose $\gamma<0$).

Turning to the rotational energy ($O(1/N_c)$) we see that in all cases it determined by the \eq{en-SU3} of the text with
moment of inertia:
\beq
I_1 =-\frac{3}{20\gamma} N_c
\eeq
and mixing coefficient $\tilde{a}_K=2/5$ both for the excitations in the sector of $u,d$-quarks, \eq{70ud-toy}, and
$s$-quarks, \eq{70s-toy}.

The moment of inertia $I_1$ does not coincide with the Inglis \cite{Inglis} moment of inertia $I_{\text{Inglis}}$ obtained in the cranking approximation:
\begin{widetext}
\beq
{\cal E}_{\text{rot}}=\frac{N_c}{2}\sum_{m=\text{free}}\frac{ <0|
\half(\bm\tOmega\cdot\bm\lambda+\bm\tomega\cdot\bm\sigma)|m><m|\half(\bm\tOmega\cdot\bm\lambda+\bm\tomega\cdot\bm\sigma)
|0>}{\varepsilon_m-\varepsilon_0}=\frac{1}{2}I_{Inglis}(\bm\tOmega-\bm\tomega)^2
, \qquad
I_{\text{Inglis}}=-\frac{1}{8\gamma} N_c
\la{Inglis-toy}
\eeq
\end{widetext}
This was noticed for the first time in \cite{Pobylitsa}. Hence in this model there is non-zero mixing of rotations with other quantum fluctuations of $\rho^a_i$ field.

To account for effect of this mixing in the  rotational energy we have to find the correction $\delta\rho^a_i$ to the mean field. We solve the Dirac equation with rotational correction and self-consistency equation:
\[
\varepsilon\phi^{\alpha i} =\rho^a_l(\lambda^a)^\alpha_\beta (\sigma^l)^i_j\phi^{\alpha' i'}
+\half \tomega_l (\sigma_l)^i_{i'}\phi^{\alpha i'}+\half \tOmega_l (\lambda_l)^\alpha_{\alpha'}\phi^{\alpha' i},
\]\beq
\qquad
\rho^a_i=\phi^+\sigma_i\lambda^a\phi
\eeq
in the leading order in spin frequency $\omega_l$ and flavor frequency $\Omega^a$ (we restrict ourselves to the
$a=1\ldots 3$ which are only relevant at the moment)  and obtain
\[
\delta\psi^{\alpha i} = -\frac{3}{20\sqrt{2}}\left(
\tomega_l(\sigma_l)^i_{i'}\varepsilon^{\alpha i'}+\tOmega_a(\lambda_a)^\alpha_{\alpha'}\varepsilon^{\alpha' i}
\right),\]\beq
\delta\rho^8_i = \frac{\sqrt{3}}{10\gamma}(\tOmega_i-\tomega_i)
\eeq
The last equation here gives the change of the mean field we are looking for. Due to the hedgehog symmetry $\delta\rho$ depends only on difference of flavor and spin frequencies. The correction to energy due to rotation is (here the first term is 2nd order expansion of quark determinant in rotation and change of mean field $\delta\rho$ and the second is  change of pure meson Lagrangian.)
\begin{widetext}
\beq
{\cal E}_{\text{rot}}=\frac{N_c}{2}\sum_{m=free}\frac{ \langle 0|\gamma\delta\rho^a_i\lambda^a\sigma_i+\half(\bm\tOmega\cdot\bm\lambda+\bm\tomega\cdot\bm\sigma)|m\rangle
\langle m|\gamma\delta\rho^a_i\lambda^a\sigma_i+\half(\bm\tOmega\cdot\bm\lambda+\bm\tomega\cdot\bm\sigma)|0\rangle}{\varepsilon_m-\varepsilon_0}+
\frac{\gamma N_c}{2}\left(\delta\rho^a_i\right)^2
=-\frac{3N_c}{40\gamma}(\bm\tOmega-\bm\tomega)^2
\eeq
\end{widetext}
which is, indeed, rotational energy with moment of inertia $I_1$.  Let us stress again that the fact
that this expression is different from the cranking approximation \eq{Inglis-toy} is completely due to the mixing of rotational degrees of freedom with
other quantum fluctuations. Let us note that this mixing is absent if the flavor group is $SU(2)$. In general, the mixing appears because the model is non-relativistic (see Appendix B).

There are two types of one-quark excitations: in the sector with $u,d$-quarks (wave function $\phi_1$) and in the sector of $s$-quarks (wave
function $\phi_s$). Mixing coefficients $a_K$ are determined by linear terms in the frequencies (see Section \ref{Theory}). We have:
\[
\delta{\cal S}_{\text{rot}}=-\int\!dt\sum_{m=\exc}\left[
\langle m|\half(\bm\tOmega\cdot\bm\lambda+\bm\tomega\cdot\bm\sigma)|m \rangle+
\right.
\]\beq
\left.
+\langle m|\delta\rho^a_i\lambda^a\sigma_i|m \rangle\right]
\equiv\! -\!\int\!\! dt \left[(1-a_K)\bm\tomega_K\cdot\bm K +a_K\tOmega_K\cdot\bm K \right]
\la{def-ak-toy}
\eeq
For $K=1$ level in the sector of $u,d$-quarks we introduce wave function of excited level as $\phi^{\alpha i}_{\exc}=\sum \chi_{K_3}\phi^{\alpha i}_{1(K_3)}$, then
\[
\langle \exc|\half(\bm\tOmega\cdot\bm\lambda+\bm\tomega\cdot\bm\sigma)|\exc\rangle=
\]\[
=\frac{1}{2}
\sum_{K_3,K_3'}\langle K'_3|(\bm\tOmega+\bm\tomega)\cdot\bm K|K_3\rangle \chi^+_{K'_3}\chi_{K_3}
\]
and
\[
\langle \exc|\delta\rho^a_i\sigma_i\lambda_a|\exc\rangle=
\]\beq
-\frac{1}{10}
\sum_{K_3,K_3'}\langle K'_3|(\bm\tOmega-\bm\tomega)\cdot\bm K|K_3\rangle \chi^+_{K'_3}\chi_{K_3}
\eeq
Comparing with the \eq{def-ak-toy} we see that in this case $\tilde{a}_{K}=\half-\frac{1}{10}=\frac{2}{5}$.
In terms of Section \ref{Theory} we can say that $c_K=0$ (this is the consequence of the fact that in our
theory there is no orbital momenta) and mixing coefficient $\zeta=1/10$.

For excitations of $s$-quark wave function is $\phi^{\alpha i}_{\exc}=\sum \chi_{j_3}\phi^{\alpha i}_{s(j_3)}$
where $j=\half$ is total momentum of s-quark excitation. Then:
\[
\langle \exc|\half(\bm\tOmega\cdot\bm\lambda+\bm\tomega\cdot\bm\sigma)|\exc\rangle=
\sum_{j_3,j_3'}\langle j'_3|\bm\tomega\cdot\bm j|j_3\rangle \chi^+_{j'_3}\chi_{j_3}
\]\beq
\langle \exc|\delta\rho^a_i\sigma_i\lambda_a|\exc\rangle=
\frac{2}{5}
\sum_{j_3,j_3'}\langle j'_3|(\bm\tOmega-\bm\tomega)\cdot\bm j|j_3\rangle \chi^+_{j'_3}\chi_{j_3}
\eeq
From the second of these expressions we obtain again the mixing coefficient $\tilde{a}_K=\frac{2}{5}$
(the role of $\bm K$ is played now by $\bm j$).

Thus the mean field approximation correctly reproduces exact formulae for energy \eq{70ud-toy} and \eq{70s-toy} in all cases.

\section{Mixing of the slow rotations in relativistic theories}

Rotational degrees of freedom can mix with other quantum fluctuations. This phenomenon is known well in nuclear
physics \cite{T-V}. As we show below this phenomenon is absent for soliton constructed within the mean field approximation (at $N_c\to\infty$) with help of relativistically invariant meson Lagrangian. This statement is true, at least, for solitons made of pions , e.g. in the Skyrme model \cite{Adkins:1983ya} or quark-soliton chiral model \cite{Diakonov:1986yh}.

Indeed, let us consider in the Skyrme model flavor rotation $R(t)$ together with some other quantum fluctuations $\delta\pi^a$.
The general pion field is presented as:
\renewcommand{\theequation}{\Alph{section}\arabic{equation}}
\beq
\pi_a(\bm x,t)\lambda_a = R(t)\left[\bar\pi^a(\bm x)+\delta\pi(\bm x,t)\right]\lambda_a R^+(t)
\la{ap2-pi}
\eeq
where $\bar\pi_a(\bm x)$ is the time independent mean field. We have to substitute \eq{ap2-pi} into the Skyrme action
or action obtained by integration over quarks in quark-soliton model and analyze the contribution of fluctuations $\delta\pi$ to effective Lagrangian.

Mixing of the rotations with other quantum fluctuations implies  terms of the form $\delta\pi(x,t){\cal K}\Omega$ where $\Omega$ is the flavor
frequency and ${\cal K}$ is some operator. The appearance of $\Omega$ means that the mixing can arise only from those terms of effective chiral
Lagrangian which contain time derivatives. However, in the relativistic theory there is usually an even number of time derivative and ar least
two of them. From the other hand, in the leading order in $N_c$ we can  consider frequency $\Omega$ to be constant in time. Therefore, the
second derivative should be applied to $\delta\pi$, so $\K$  is at least linear in time derivatives. However, all such terms are full
derivatives and can be omitted.

The noticable exclusion from this rule is Wess-Zumino-Witten term which is linear in time derivative. We have to apply
this derivative to $R(t)$ in order to obtain flavor frequency. Using independence of $\Omega$ on coordinates we
arrive at (see, e.g., \cite{Diakonov:1986yh}):
\[
\delta L = \frac{N_c}{48\sqrt{3}\pi^2}\int\! dt\; \Omega_8\int\! d^3x\; \varepsilon_{ijk}\times
\]\beq
\times \Tr\left[\lambda_8 \left( (U^+\partial_i U\right)\left(U^+\partial_j U\right)\left(U^+\partial_k U\right) \right] \eeq where $U=\exp
i(\bar\pi^a+\delta\pi^a(x,t))\lambda^a$ is the pion mean field together with quantum fluctuations. The quantity: \beq Q_t =\frac{1}{24\pi}\int\!
d^3x\; \varepsilon_{ijk}\Tr\left[ \lambda_8 \left(U^+\partial_i U\right)\left(U^+\partial_j U\right)\left(U^+\partial_k U\right)\right] \eeq is
topological charge of the field $U$ and cannot be changed by the small fluctuations of the pion field. In other words: \beq \frac{\delta
Q}{\delta\pi(x,t)} = 0 \eeq Hence mixing of rotations with quantum fluctuations is absent in the Skyrme model.

Situation is similar in the quark-soliton model \cite{Diakonov:1986yh}. Mixing can appear only from so-called
"imaginary part" of the effective $\pi$-meson action. This part of the action starts by WZW term but in principle
is an infinite series in gradients of pion field. However all these terms are full derivatives and the sum reduce to
the complete baryon charge of the state. This quantity is determined by the number of valence quarks and cannot be changed by the small fluctuations of the pion field.

This does not mean that mixing is always zero. First, it can appear in more general meson Lagrangians. Second, it
can arise if the frequencies of rotations are not small. For example, properties of rotational exotic states
(at $\Omega_{4,5,6,7}\sim O(1)$) can be described as mixing of rotations and quantum $K$-meson states belonging to the
continuum spectrum. This leads to the width of these states $~\sim O(1)$. However, the rotational theory in this
case should be modified anyway, as due to the WZW term rotations in the strange directions turn into small oscillations (see, e.g.
\cite{Callan-Klebanov,kleb2} and \cite{Diakonov:2008hc}).

\appendix

\setcounter{section}{2}%

\section{Matrix elements of one-particle operators}
Dirac equation in the sector of $u,d$-quarks conserves the grand spin $K$. The angular part of the wave function of
the state with given $K$ is spherical spinor-isospinor:
\beq
\Xi^{\alpha i}_{KK_3jl}(\bm n)=\sum_{j_3} C^{KK_3}_{jj_3;\half\alpha}\Omega^i_{jj_3l}(\bm n)
\eeq
Here $\alpha$ is isospinor and $i$ is spinor indices, $\Omega$ is a spherical spinor with total angular moment $j$ (projection $j_3$) and orbital momentum $l$; Clebsh-Gordan coefficient $C^{\ldots}_{\ldots}$ joins $j$ and isospin $\bm t$
($t=\half$) into the grand spin $\bm K$. Total angular momentum can be $j=K\pm \half$. Spherical spinor $\Omega^i$
is constructed according to the same rule out of spin of the quark $\bm s$ ($s=\half$) and orbital momentum $\bm l$
\beq
\Omega^{i}_{jj_3l}(\bm n)=\sum_{j_3} C^{jj_3}_{ll_3;\half i}Y_{ll_3}(\bm n)
\eeq
where $Y_{ll_3}(\bm n)$ are usual spherical harmonics.

We are looking for the solution of the Dirac equation: $\varepsilon\Psi={\cal H}\Psi$ with the Dirac Hamiltonian
\eq{DiracH} which is a bispinor $\{\varphi,\chi\}$ in the form:
\beq
\Psi^{\alpha i}= \left(
\begin{array}{c}
g(r)\Xi^{\alpha i}_{KK_3,K-\half,K}+h(r)\Xi^{\alpha i}_{KK_3,K+\half,K}\\
f(r)\Xi^{\alpha i}_{KK_3,K-\half,K-1}+j(r)\Xi^{\alpha i}_{KK_3,K+\half,K+1}\\
\end{array}
\right)
\eeq
This state has a "natural parity" $(-1)^K$. Indeed, parity transformation is $\varphi(\bm r)\rightarrow \varphi(-\bm r)$, $\chi(\bm r)\rightarrow -\chi(-\bm r)$ and hence parity is determined by the value $l$. The state with parity
$(-1)^{K+1}$ corresponds by exchange of $\varphi$ and $\chi$ in this expression. At $K=0$ wave function is determined only by 2 functions $h(r)$ and $j(r)$.

We want to calculate the matrix elements of $\bm t$. Since it acts only on isospinor indices $\alpha$ and spherical spinors $\Omega$ are orthonormal, we obtain:
\[
\langle K'_3|\bm t|K_3\rangle = \sum_{\alpha\beta,j,j_3} \left(C^{KK'_3}_{jj_3\half\alpha}\right)^*\langle\alpha|\bm
t|\beta\rangle C^{KK_3}_{jj_3\half\beta}\times
\]\beq
 \left[(1-c_K)\delta_{j K-\half}+c_K\delta_{j K+\half}
\right]
\eeq
with $c_K$ defined by \eq{toapp} and $\langle\alpha|\bm t|\beta\rangle$ are usual generators of isospin $\half$.

Substituting Clebsh-Gordan coeffients we arrive at \eq{mat-K}. In particular, if $\bm t \rightarrow t_3$ the matrix elements are diagonal in $\alpha, \beta$ and hence diagonal in $K_3,K'_3$. The matrix element:
\[
\langle K_3|t_3|K_3\rangle = \half \left(
\left|C^{KK_3}_{j,K_3-\half;\half,\half}\right|^2-\left|C^{KK_3}_{j,K_3+\half;\half,-\half}\right|^2
\right)\times
\]\beq
\times\left[(1-c_K)\delta_{j K-\half}+c_K\delta_{j K+\half}
\right]
\eeq
and using expressions for  Clebsch-Gordan coefficients we obtain
\beq
\langle K_3|t_3|K_3\rangle=K_3\left(\frac{1-c_K}{2K}-\frac{c_K}{2(1+K)}\right)
\eeq
It is a particular case of \eq{mat-K}.

\section{Decays of excited baryons}
Calculations of the widths of excited baryons is outside the scope of this paper, however, in this appendix we present only general discussion of the baryon decays. For the ground state baryons the procedure of calculation is known: was constructed for the Skyrme model already in \cite{Adkins:1983ya}, for the quark-soliton model see, e.g. \cite{Diakonov:1986yh,Blotz,Rev}. At last, strictly in the limit $N_c\to\infty$ decay constants in the approach \cite{DJM1} were calculated, e.g., in \cite{carson,Goity2} and in other, already cited papers.

Typical decays of excited baryons below 2 GeV are of the type $B_i\to B_fM$ with one emitted meson, at least such decays always give essential part of the width. To be specific, we will talk about decays into $\pi$-mesons. Let us estimate the width in the limit of large $N_c$. (This estimate is already known: we are close here to  J.L.Goity in \cite{Goity:2005fj} (see also \cite{Goity2})).  The one pion decays of the excited baryons are described by the effective Lagrangian of the type:
\beq
{\cal L}_{eff}=\frac{g_a}{F_\pi}\int\! d^3x \;\bar{\Psi}^{(f)}_B\gamma_\mu\gamma_5\frac{\lambda^a}{2}\Psi^{(i)}_B \partial_\mu\pi
\eeq
Here   $\Psi^{(i)}_B$ and $\Psi^{(f)}_B$ are fields of initial and final baryon, $\pi$ is the $\pi$-meson field with flavor $a$, $\lambda_a$ is the corresponding Gell-Mann matrix. At last $g_a$ is the  transitional axial coupling constant. The width $\Gamma_{fi}$ of partial decay to $B_f\pi$ is proportional to coupling constant squared and phase volume:
\beq
\Gamma_{fi} \sim \frac{g^2_a}{8\pi F^2_\pi}\Delta^3
\la{width}
\eeq
(see, e.g. \cite{Diakonov:2008hc}) where $\Delta=M_{i}-M_f$ is the difference of mass of the initial and final baryon.

The coupling constant can be calculated as a matrix element of the corresponding quark operator between mean field initial and final state:
\beq
g_a(k)\sim \int\! d^3x\; \langle {\rm\small fin}|\bar{\psi}\gamma_5\gamma_\mu\psi(x)|{\rm\small in}\rangle e^{ikx}
\la{coupling}
\eeq
The role of quark operator is played by axial current for decays with $\pi$-mesons, vector current for decays into $\rho$-mesons, etc.
Expression \eq{coupling} already implies the $N_c\to\infty$ limit, as baryons are considered to be  heavy (mass $O(N_c)$) non-relativistic objects. (Expression \eq{width} is also written in this limit). Plane wave $e^{ikx}$ represents wave function of emitted meson, with $k$ being its momentum. At last $|{\rm in}\rangle$ and $|{\rm fin}\rangle$ are mean field approximations for initial and final baryon quark wave functions. They are product of all 1-quark wave functions --- solutions of Dirac equation in the mean field --- for all filled levels. In general, one has to write here wave functions rotated by matrices $R$ and $S$ in order to take into account degeneracy of the mean field. After the calculation of matrix element \eq{coupling} we obtain some operator depending on collective coordinates. Averaging this operator with collective wave functions of initial and final baryon we obtain the coupling constant for some specific decay.

In fact, \eq{coupling} is only the first term of expansion in the time derivatives of collective coordinates. Next terms can be obtained in the same manner as it was done for corrections in $m_s$ in the main text. Due to the limit $N_c\to\infty$ all collective coordinates are slowly varying functions of coordinates, so the expansion
in time derivatives is an expansion in $1/N_c$, with \eq{coupling} being its leading term. For ground state baryons and for decays into pions the corresponding formulae were presented in \cite{Rev}.

Decays of excited baryons are possible either to the baryons belonging to the same rotational band or to the baryons which have the different filling of intrinsic quark levels (e.g., to ground state baryons). In the first case the coupling constant is large, $O(N_c)$. Example is  the transitional axial constant $g_a(\Delta\to N\pi)$ \cite{Adkins:1983ya}. In the second case the coupling constant is always smaller. This difference is clearly seen from \eq{coupling}.

Indeed, for the configuration of levels being the same for the initial and final state, the coupling constant is a sum of $N_c$ 1-particle matrix elements corresponding  to all $N_c$ quarks. If excited quark changes its intrinsic state then only one of $N_c$ contributions would survive, which is the overlap 1-particle matrix element between initial and final states of this quark (all other are zero due to orthogonality of wave functions). However, if the final state is the ground state additional factor $\sqrt{N_c}$ appears which is due to the different normalization of initial and final wave functions:
\[
g_a(R,S) \sim \int d^3x\! \phi^*_f({\bf x})S^+\gamma_3\gamma_5SR^+\frac{\lambda^a}{2}R\phi_i({\bf x})j_l(k|{\bf x}|)\times
\]\beq
\times
{\cal D}^l_{m_1,m_2}(S)
Y_{lm_2}\left(\frac{{\bf x}}{|{\bf x}|}\right)\!
\left\{
\begin{array}{cc}
N_c & i,f = {\rm same\; band}\\
\sqrt{N_c} & i={\rm excited},f\! =\! {\rm ground} \\
1 & i={\rm excited},f\!=\! {\rm excited'}
\end{array}
\right.
\la{axial}
\eeq
This expression is written a bit schematically.
Wave functions $\psi_i$ and $\psi_f$ are initial and final wave functions of excited quark, $R$ is a rotational matrix in flavor and $S$ in ordinary space, ${\cal D}^l_{m_1m_2}(S)$ is Wigner function and $Y_{lm}$ is an ordinary spherical harmonics (summation in all possible $m_2$ is implied). At last $j_l(kr)$ is a spherical Bessel function. It appears (together with spherical harmonics) as a result of expansion of a plane wave in \eq{coupling} into the set of spherical waves. If the momentum of emitted meson is small $ka\ll 1$ ($a$ is the scale of wave functions $\psi_{i,f}$ which coincides with the characteristic size of the baryon) it is sufficient to account only for the lowest angular momentum $l=0$ (angular momentum of emitted pion is 1).

Axial constant \eq{axial} is an operator in the space of collective coordinates (derived in the leading order in $N_c$). To obtain coupling constant responsible for decay of concrete baryon to another one, we have to average expression \eq{axial} with collective wave functions:
\beq
g_a(i\to f)= \int\! dR dS\; \psi^{(rot)*}_f(R,S)g_{a}(R,S)\psi^{(rot)}_i(R,S)
\la{if-axial}
\eeq

\begin{center}
\begin{figure}[htb]
\includegraphics[width=5 cm]{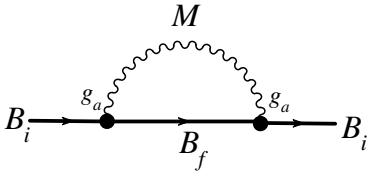}
\caption{Self energy correction to the excited baryon mass}
\label{fig:5}
\end{figure}
\end{center}

In spite of the fact that coupling constants are smaller, the widths of decays to the different quark levels are typically larger in $N_c$. The reason is that phase volume in this case is always larger. The mass differences are $O(1/N_c)$ for decays inside the same rotational band but $O(1)$ for transitions with change intrinsic state of excited quark. As a result the widths of decays inside the rotational band are suppressed as $O(1/N^2_c)$ while decays of excited baryons with discharge of the excitation are always $O(1)$ (and decays to the the other levels are suppressed). In particular, the total width of ground state baryons (decuplet with spin $\frac{3}{2}$ ) is only $O(1/N^2_c)$, while all remaining baryons have total width of $O(1)$.

In practical terms only decays to the ground octet or decuplet are observable. For all baryons they have partial widths independent on $N_c$ up to corrections in $1/N_c$ which can be still essential at $N_c=3$. Let us prove the theorem: widths of all baryons belonging to the same rotational band are the same in the leading order in $N_c$.

Indeed, mass differences of all baryons entering the same rotational band are the same in the leading order in $N_c$. Hence
\beq
\Gamma^{(i)}_{tot}=\sum_f \Gamma(i\to f) = \frac{\Delta^3}{8\pi}\sum_f g^2_a(i\to f) = \Gamma_{level}
\eeq
However, the sum of axial constants squared over all possible final states does not depend on the initial state of the band. According to \eq{if-axial} axial constant squared contains two integrals in $R,S$ and  $R',S'$. Due to completeness of final baryon rotational functions:
\[
\sum_f  \psi^{(rot)*}_f(R,S)\psi^{(rot)}_f(R',S') = \delta(R-R')\delta(S-S')
\]
leads to $R=R'$ and $S=S'$. Then the sum in all possible flavors of pseudoscalar mesons and directions of axial current gives the expression which does not depend on $R$ and $S$ due to Fiertz identities. Dependence on matrices remains only in the initial wave function. Integral over $R$ and $S$ is becoming the normalizing integral for initial collective wave function and the dependence on initial state disappears completely. Obtained total width has a sense of the complete width of the intrinsic quark level and is universal for the whole rotational band around it.

The proved theorem is broken strongly in nature. There are many reasons for that: corrections in $N_c$ and mass of the strange quark $m_s$ to
the coupling constants, mixing of multiplets, etc. Perhaps, the strongest source of the deviations is simply the difference in the phase volumes
(which is $O(1/N_c)$ effect) for different baryons entering the same rotational band.

The fact that widths of excited baryons are not suppressed in the large $N_c$ limit, as it was mentioned in the Introduction, makes these baryons not well-defined. One can count only on numerical smallness of the width not related to $N_c$. In such a situation baryon resonances can be defined only as poles in the complex plane of meson-nucleon scattering amplitude. This approach was applied in  \cite{Diakonov:2008hc} to the problem of pentaquark (which also has width independent on $N_c$) for the Skyrme model but in general case it looks too complicated. If the width is small
one returns to the self-consistent field description presented here.

It seems  that width of the baryon which is not suppressed at $N_c\to\infty$ possesses the danger to our approach in general. Indeed, due to unitarity non-zero width implies not only imaginary part of the pole but also a shift in  the real part, i.e. leads to the change of the baryon mass. It can be small numerically but it is $O(1)$ in $N_c$. If it is different for the baryons entering the same rotational band, our formulae for mass splittings inside the rotational band would become senseless. Fortunately, it is not the case.

Corrections to the mass  due to decays into the $\pi$-mesons are presented by self-energy diagram Fig.\ref{fig:5}. The imaginary part of this diagram gives the width of $B_i\to B_f M$ decay, real part gives the shift of mass. The point is that the mass shift does not depend on the baryon $B_i$ on the same rotational
band:
\[
\Delta M \sim \sum_f\int\frac{d^4k}{(2\pi)^4}\frac{g^2_a(i\to f)}{k^2(\Delta+k_0)}
\]
as it was proved above. Hence we arrive at conclusion that mass shifts are universal for all baryons inside the rotational bands. It can be included to the general shift of the intrinsic level  and does not break down the mass relations in the $O(1/N_c)$ order derived in the text. Next order corrections in $N_c$ due to the finite width of the resonance also do not destroy these relations. However, they can renormalize the moment of inertia $I_1$. Example of
such a situation is given by pentaquark in the Skyrme model \cite{Diakonov:2008hc}.

\section{Isoscalar factors for $N_c\to\infty$\la{Cle-Gor}}
Clebsch-Gordan coefficients for large $N_c$-baryon multiplets were calculated in a number of References\cite{Pra,Kaplan,DJM1}. Two methods can
be used for this calculation: either based on decomposition $SU(3)$ spinor with large number of indices\cite{Kaplan,DJM1} or applying lowering
and rising generators in the given representation  to the state with highest weight\cite{Pra}.

However, the tables of CG-coefficients in  above references are usually incomplete and do not correspond to the conventions of the
\cite{deswart} (which serves as  a common standard for $SU(3)$-group) but differ from this standard by unitary transformation. This is
inconvenient and for this reason we give here the complete tables for isoscalar factors at arbitrary $N_c$.

During the refereeing we have been informed that the complete tables of isoscalar coefficients were presented in \cite{Cohen:2004ki}. The tables
presented here are analogous and differs only method of calculation. Nonetheless, we leave here the tables for the completeness of description.
We thank referee for taking our attention for this article.

\begin{center}
\begin{table*}
\begin{tabular}{|c|c|c|c|c|c|c|}
\hline
 & $\eta N\to N$ &  $\pi N\to N$ & $K\Sigma\to N$ & $K\Lambda\to N$ & $\bar{K}N\to \Sigma$
 &$\eta\Sigma\to\Sigma$\\
 \hline
\quad F$\:\:\:\:$ &$-\frac{3(\nu^2+3\nu+1)}{\sqrt{2(\nu+2)(\nu+4)}}$ &
$\frac{\nu^2+5\nu+9}{\sqrt{2(\nu+2)(\nu+4)}}$ &
$\frac{(11+4\nu)\sqrt{\nu}}{\sqrt{2(\nu+2)(\nu+4))}}$ &
$\frac{\sqrt{3}(2\nu+3)}{\sqrt{2(\nu+4)}}$
& $-\frac{\sqrt{\nu}(11+4\nu)}{\sqrt{3(\nu+2)(\nu+4)}}$
& $\frac{(1-\nu)(8+3\nu)}{\sqrt{3(\nu+2)(\nu+4)}}$
\\
\hline
\quad D$\:\:\:\:$& $\sqrt{\frac{\nu}{2}(\nu+2)}$ &
$ -3\sqrt{\frac{\nu}{2}(\nu+2)}$ &
$3\sqrt{\frac{\nu}{2}+1}$ & $-\sqrt{\frac{3\nu}{2}}$& $-\sqrt{3(2+\nu)}$
& $(3-\nu)\sqrt{\frac{2\nu+3}{2\nu}}$\\
\hline
& $\pi\Sigma\to\Sigma$ & $\pi\Lambda\to\Sigma$ & $K\Xi\to\Sigma$ & $\bar{K}N\to\Lambda$ & $\eta\Lambda\to\Lambda$ & $\pi\Sigma\to\Lambda$
\\
\hline
\quad F$\:\:\:\:$ &
$\frac{\sqrt{\nu+4}(\nu+5)}{\sqrt{3(\nu+2)}}$ &
$\frac{(1-\nu)\sqrt{\nu}}{\sqrt{6(\nu+4)}}$ &
$\frac{7\nu+8}{3\sqrt{\nu+4}}$ &
$\frac{\sqrt{3}(2\nu+3)}{\sqrt{\nu+4}}$ &
$\frac{3(2-\nu-\nu^2)}{\sqrt{2(\nu+2)(\nu+4)}}$ &
$\frac{(\nu-1)\sqrt{\nu}}{\sqrt{2(\nu+4)}}$
\\
\hline
\quad D$\:\:\:\:$ &
$(1-\nu)\sqrt{\frac{3(2+\nu)}{\nu}}$ &
$\sqrt{\frac{3}{2}}(\nu+1)$ &
$-\frac{2\nu+1}{\sqrt{\nu}}$ &
$-\sqrt{3\nu}$ &
$-\frac{(\nu+5)\sqrt{\nu}}{\sqrt{2(\nu+2)}}$ &
$-\sqrt{\frac{3}{2}}(\nu+1)$
\\
\hline
 &$K\Xi\to\Lambda$
&$\bar{K}\Sigma\to\Xi$ & $\bar{K}\Lambda\to\Xi$ &$\eta\Xi\to\Xi$ & $\pi\Xi\to\Xi$ &\\
\hline
\quad F$\:\:\:\:$ &
$\frac{5\sqrt{\nu(\nu+2)}}{\sqrt{\nu+4}}$ &
$\frac{7\nu+8}{\sqrt{6(\nu+4)}}$ &
$-\frac{5\sqrt{\nu(\nu+2)}}{\sqrt{2(\nu+4)}}$ &
$\frac{(8-3\nu)\sqrt{\nu+2}}{2\sqrt{\nu+4}}$ &
$\frac{(16-\nu)\sqrt{\nu+2})}{3\sqrt{2(\nu+4)}}$  &
 \\
\hline
\quad D$\:\:\:\:$ &
$\frac{3}{\sqrt{\nu+2}}$ &
$-\frac{(2\nu+1)\sqrt{3}}{\sqrt{2\nu}}$ &
$\frac{3}{\sqrt{2(\nu+2)}}$ &
$\frac{3-5\nu-\nu^2}{\sqrt{2\nu(\nu+2)}}$ &
$\frac{\nu^2+3\nu+5}{\sqrt{2\nu(\nu+2)}}$ &
\\
\hline
\end{tabular}
\caption{Isoscalar factors for $8_M\otimes "8"_B\to "8"_B$ at arbitrary $N_c=2\nu+1$. In the table two isoscalar factor for antisymmetrical (F) and symmetrical (D) case are presented. Any value from the table should be divided by universal factor $\sqrt{5\nu^2+16\nu+9}$ which we omit for brevity. Definition of isoscalar factors correspond to conventions of \cite{deswart} and reduce to the usual ones at $N_c=3$}
\end{table*}
\end{center}

\begin{center}
\begin{table*}
\begin{tabular}{|c|c|c|c|c|c|c|}
\hline
 & $K\Sigma^*\to\Delta$ &  $\eta\Delta\to\Delta$ & $\pi\Delta\to\Delta$ & $K\Xi^*\to\Sigma^*$ & $\pi\Sigma^*\to \Sigma$
 &$\eta\Sigma^*\to\Sigma^*$\\
 \hline
\quad F$\:\:\:\:$ & $\frac{\sqrt{5(\nu+4)}(3\nu+2)}{2\sqrt{(\nu+2)(\nu+6)}}$ & $-\frac{\sqrt{5}(\nu^2+5\nu+3)}{\sqrt{2(\nu+2)(\nu+6)}}$ &
$\frac{\nu^2+7\nu+15}{\sqrt{2(\nu+2)(\nu+6)}}$ & $\frac{2\sqrt{5(\nu+3)}(2\nu+3)}{3\sqrt{(\nu+2)(\nu+6)}}$ &
$\frac{\sqrt{5}(\nu^2+5\nu+12)}{2\sqrt{3(\nu+2)(\nu+6)}}$ & $-\frac{\sqrt{5}\nu(\nu+3)}{\sqrt{2(\nu+2)(\nu+6)}}$
\\
\hline
\quad D$\:\:\:\:$&
$-\frac{\sqrt{\nu}}{2}$ &
$-\sqrt{\frac{\nu(\nu+4)}{2}} $&
$\sqrt{\frac{5\nu(\nu+4)}{2}}$ &
$-\frac{2\sqrt{\nu(\nu+3)}}{3\sqrt{\nu+4}}$ &
$-\frac{(19+5\nu)\sqrt{\nu}}{2\sqrt{\nu+4}}$ &
$-\frac{(\nu+5)\sqrt{\nu}}{\sqrt{2(\nu+4)}} $
\\
\hline
& $\bar{K}\Delta\to\Sigma^*$ & $\pi\Xi^*\to\Xi^*$ & $\eta\Xi^*\to\Xi^*$ & $K\Omega\to\Xi^*$ & $\bar{K}\Sigma^*\to\Xi^*$ & $\bar{K}\Xi^*\to\Omega$
\\
\hline
\quad F$\:\:\:\:$ &
$\frac{\sqrt{5(\nu+4)}(2\nu+3)}{\sqrt{3(\nu+2)(\nu+6)}}$ &
$\frac{\sqrt{5}(\nu^2+3\nu+9)}{3\sqrt{2(\nu+2)(\nu+6)}}$ &
$\frac{\sqrt{5}(3-\nu-\nu^2)}{\sqrt{2(\nu+2)(\nu+6)}}$ &
$\frac{\sqrt{5}(2\nu+3)}{\sqrt{2(\nu+6)}}$ &
$\frac{\sqrt{10(\nu+3)}(2\nu+3)}{\sqrt{3(\nu+2)(\nu+6)}}$ &
$\frac{\sqrt{5}(2\nu+3)}{\sqrt{\nu+6}}$
\\
\hline
\quad D$\:\:\:\:$ &
$-\sqrt{\frac{\nu}{3}}$ &
$-\frac{(5\nu+18)\sqrt{\nu}}{3\sqrt{2(\nu+4)}}$ &
$-\frac{(\nu+6)\sqrt{\nu}}{\sqrt{2(\nu+4)}}$ &
$-\sqrt{\frac{\nu(\nu+2)}{2(\nu+4)}} $ &
$-\sqrt{\frac{2\nu(\nu+3)}{3(\nu+4)}}$ &
$-\sqrt{\frac{\nu(\nu+2)}{\nu+4}}$
\\
\hline
 &$\eta\Omega\to\Omega$&&&&&
\\
\hline
\quad F$\:\:\:\:$ &
$\frac{(3-\nu)\sqrt{5(\nu+2)}}{\sqrt{2(\nu+6)}}$&&&&&
\\

\hline
\quad D$\:\:\:\:$ &
$-\frac{(\nu+7)\sqrt{\nu}}{\sqrt{2(\nu+4)}}$
&&&&&
\\
\hline
\end{tabular}
\caption{Isoscalar factors for $8_M\otimes "10"_B\to "10"_B$ at arbitrary $N_c=2\nu+3$. Two factors for antisymmetrical (F) and symmetrical (D) case
are given. The table value  should be divided by  factor $\sqrt{3\nu^2+16\nu+15}$. Definition of isoscalar factors correspond to conventions of \cite{deswart} and reduce to the usual ones at $N_c=3$. In particular, at $N_c=3$ only symmetrical representation survives}
\end{table*}
\end{center}

Consulting with tables one can see that the change of Clebsch-Gordan coefficients from $N_c\to\infty$ to $N_c=3$ is, indeed, rather large. In particular, one can note the cases when isoscalar factor changes the sign during this transition. Moreover, for "decuplet" there are {\em two} possible
final multiplets at $N_c\ge 5$, so one can discuss $F/D$ ratio for "decuplet". Corresponding multiplet dies out as $N_c=3$.

From the other hand, Clebsch-Gordan coefficients can be easily taken into account at any $N_c$ and this source of inaccuracy can be avoided. For this
reason we prefer to use isoscalar factors at $N_c=3$.

\end{document}